\documentclass[a4paper]{jpconf}
\usepackage{graphicx}
\usepackage{xspace}     % Include xspace
\usepackage[usenames,dvipsnames]{color}%mjt-this uses familiar color names
\usepackage{hyperref}
\newcommand{\pT}{\mbox{$p_T$}\xspace}

\newcommand{\raa}{\mbox{$R_{\rm AA}$}\xspace}

\newcommand{\Npart}{\mbox{$N_{\rm part}$}\xspace}

\newcommand{\Et}{\mbox{${\rm E}_T$}\xspace}

\newcommand{\sqs}{\mbox{$\sqrt{s}$}\xspace}
\newcommand{\sqsn}{\mbox{$\sqrt{s_{_{NN}}}$}\xspace}
\def\lsim{\raise0.3ex\hbox{$<$\kern-0.75em\raise-1.1ex\hbox{$\sim$}}}
\def\gsim{\raise0.3ex\hbox{$>$\kern-0.75em\raise-1.1ex\hbox{$\sim$}}}

\def\mean#1{\left<#1\right>}
\def\QGP{{\color{Red} Q}{\color{Blue} G}{\color{Green} P}}
\def\QCD{{\color{Red} Q}{\color{Green} C}{\color{Blue} D}}
\begin{document}
\title{How do quarks and gluons lose energy in the \QGP?}

	\author{M.~J.~Tannenbaum\footnote{Supported by the U.S. Department of Energy, Contract No. DE-AC02-98CH10886.}}

\address{Physics Department, Brookhaven National Laboratory, Upton, NY 11973-5000 USA}

\ead{mjt@bnl.gov}

\begin{abstract}
RHIC introduced the method of hard scattering of partons as an in-situ probe of the the medium produced in A+A collisions. 
 A suppression, $\raa\approx 0.2$ relative to binary-scaling, was discovered for $\pi^0$ production in the range 
$5\leq p_T\leq 20$ GeV/c 
in central Au+Au collisions at $\sqrt{s_{NN}}=200$ GeV, and surprisingly also for single-electrons from the decay of heavy quarks. Both these results have been confirmed in Pb+Pb collisions at the LHC at $\sqrt{s_{NN}}=2.76$ TeV. Interestingly, in this $p_T$ range the LHC results for pions nearly overlap the RHIC results. Thus, due to the flatter spectrum, the energy loss in the medium at LHC in this $p_T$ range must be $\sim 40$\% larger than at RHIC. Unique at the LHC are the beautiful measurements of the fractional transverse momentum imbalance $1-\mean{\hat{p}_{T_2}/\hat{p}_{T_1}}$ of di-jets in Pb+Pb collisions. At the Utrecht meeting in 2011, I corrected for the fractional imbalance of di-jets with the same cuts in p-p collisions and showed that the relative fractional jet imbalance in Pb+Pb/p-p is $\approx 15\%$ for jets with $120\leq\hat{p}_{T_1}\leq 360$ GeV/c. CMS later confirmed this much smaller imbalance compared to the same quantity derived from two-particle correlations of di-jet fragments at RHIC corresponding to jet $\hat{p}_T\approx 10-20$ GeV/c, which appear to show a much larger fractional jet imbalance $\approx 45\%$ in this lower $\hat{p}_T$ range. The variation of apparent energy loss in the medium as a function of both $p_T$ and $\sqsn$ is striking and presents a challenge to both theory and experiment for improved understanding. There are many other such unresolved issues, for instance, the absence of  evidence for a $\hat{q}$ effect, due to momentum transferred to the medium by outgoing partons, which would widen the away-side di-jet and di-hadron correlations in a similar fashion as the $k_T$-effect. Another issue well known from experiments at the CERN ISR, SpS and SpS collider is that parton-parton hard-collisions make negligible contribution to  multiplicity or transverse energy production in p-p collisions---soft particles, with $p_T\leq 2$ GeV/c, predominate. Thus an apparent hard scattering component for A+A multiplicity distributions based on a popular formula, ${dN_{\rm ch}^{AA}/d\eta}= [(1-x) \mean{N_{\rm part}} {dN_{\rm ch}^{pp}/d\eta}/2 + x\, \mean{N_{\rm coll}} {dN_{\rm ch}^{pp}/d\eta}]$, seems to be an unphysical way to understand the deviation from $N_{\rm part}$ scaling. Based on recent p-p and d+A measurements, a more physical way is presented along with several other stimulating results and ideas from recent d+Au (p+Pb) measurements. %{\color{Red}\bf Stopped \today\ }
\end{abstract}

\section{Introduction} 
\begin{figure}[!h] 
      \centering
       \includegraphics[width=0.75\linewidth]{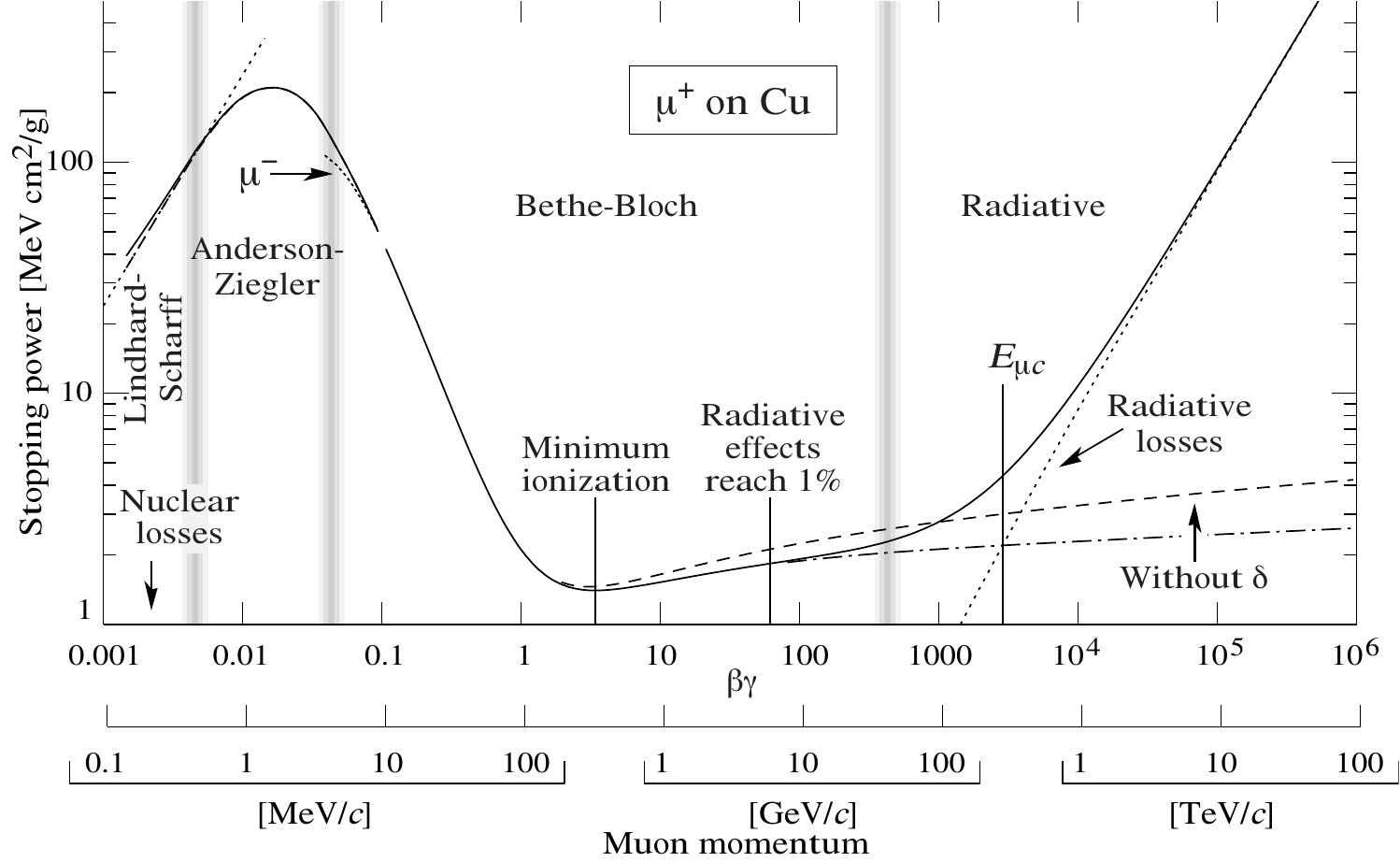}
\caption[] {$dE/dx$ of a $\mu^+$ in Copper as a function of muon momentum~\cite{rpp2008}}
      \label{fig:mudEdx}
   \end{figure}

There is a great textbook from 1950 titled \href{http://www.amazon.com/Nuclear-Physics-Course-University-Chicago/dp/0226243656}
 {``Nuclear Physics''} from a course given by Enrico Fermi~\cite{FermiNPBook}. 
Chapter I of ten is  ``PROPERTIES OF NUCLEI'' and Chapter II, ``INTERACTION OF RADIATION WITH MATTER'', which contains three sections: A) Energy Loss by Charged Particles; B. Scattering; C. Passage of Electromagnetic Radiation through Matter. Imagine that today we were going to write a textbook on \QGP\ physics. After 13 years at RHIC and 2 years at LHC-IONS, we could do a fair job on Chapter I, ``PROPERTIES OF THE \QGP''; but for Chapter  II ``INTERACTION OF QUARKS AND GLUONS AND OTHER RADIATION WITH THE \QGP", I don't know. I have not seen evidence that provides convincing proof of any theory or model. I don't know any formula comparable to Bethe-Bloch for ionization loss or Bethe-Heitler for radiation. For instance, I would like to see something for quarks and gluons in the \QGP\ like  Fig.~\ref{fig:mudEdx} for a muon in Cu. 
      %\vspace*{-0.05in}
Of course the situation is much more complicated for the \QGP\ because it is not simply a block of material but an expanding and flowing hot medium. 

There are many different theoretical studies of energy loss of a quark or gluon with their color charges fully exposed passing through a medium with a large density of similarly exposed color charges (i.e. a \QGP ). The approaches are different, but the one thing that they have in common~\cite{JETcollabarXiv1312} is the transport coefficient of a gluon in the medium, denoted $\hat{q}$, which is defined from the mean 4-momentum transfer$^2$/collision but is expressed as the mean 4-momentum transfer$^2$ per mean free path of a gluon in the medium. Thus the mean 4-momentum transfer$^2$ for a gluon traversing length $L$ in the medium is, 
$\mean{q^2(L)}=\hat{q}\,L=\mu^2\,L/\lambda_{\rm mfp}$, where $\lambda_{\rm mfp}$ is the mean free path for a gluon interaction in the medium, and $\mu$, the mean momentum transfer per collision, is the Debye screening mass acquired by gluons in the medium. In this, the original BDMPSZ formalism~\cite{BDMPSZ}, the energy loss of an outgoing parton 
due to coherent gluon bremsstrahlung per unit length ($x$) of the medium, $-dE/dx$, takes the form~\cite{BaierQM02}:
\begin{equation}
{-dE \over dx }\simeq \alpha_s \mean{q^2(L)}=\alpha_s\, \hat{q}\, L=\alpha_s\, 
\mu^2\, L/\lambda_{\rm mfp} \qquad ,
\end {equation}
so that the total energy loss in the medium goes like $L^2$~\cite{BDMPS2}. Also the accumulated transverse momentum$^2$, $\mean{k_{\perp}^2}$, for a gluon traversing a length $L$ in the medium is well approximated by $\mean{k_{\perp}^2}\approx\mean{q^2(L)}=\hat{q}\, L$.

Experiments at RHIC were the first to use hard-scattering as an in-situ probe of gluons and quarks traversing the medium in Relativistic Heavy Ion (RHI) collisions. Many important and beautiful results have been obtained using single inclusive measurements and two-particle correlations. 
A simple estimate shows that the $\mean{k_{\perp}^2}\approx\hat{q}\,L$ should be observable at RHIC via the broadening of di-hadron azimuthal correlations.  Assume that for a trigger particle with $p_{T_t}$ the away-parton traverses slightly more than half the  14 fm diameter medium for central collisions of Au+Au, say 8 fm. With a $\hat{q}=1$ GeV$^2$/fm~\cite{JETcollabarXiv1312}, this would correspond to $\mean{k_{\perp}^2}=\hat{q}\,L=8$ GeV$^2$, compared to the measured~\cite{ppg029} $\mean{k_T^2}=8.0\pm 0.2$ (GeV/c)$^2$ for di-hadrons in p-p collisions\footnote{In both cases the azimuthal projection is only half the quoted $\mean{k_T^2}$ or $\mean{k_{\perp}^2}$}  with roughly the same $p_{T_t}$ and $p_T^{\rm assoc}$. This should be visible as an azimuthal width $\sim\sqrt{2}$ larger in Au+Au than in p-p collisions at \sqsn=200 GeV. 

\begin{figure}[!h] 
      \centering
       \includegraphics[width=0.5\linewidth]{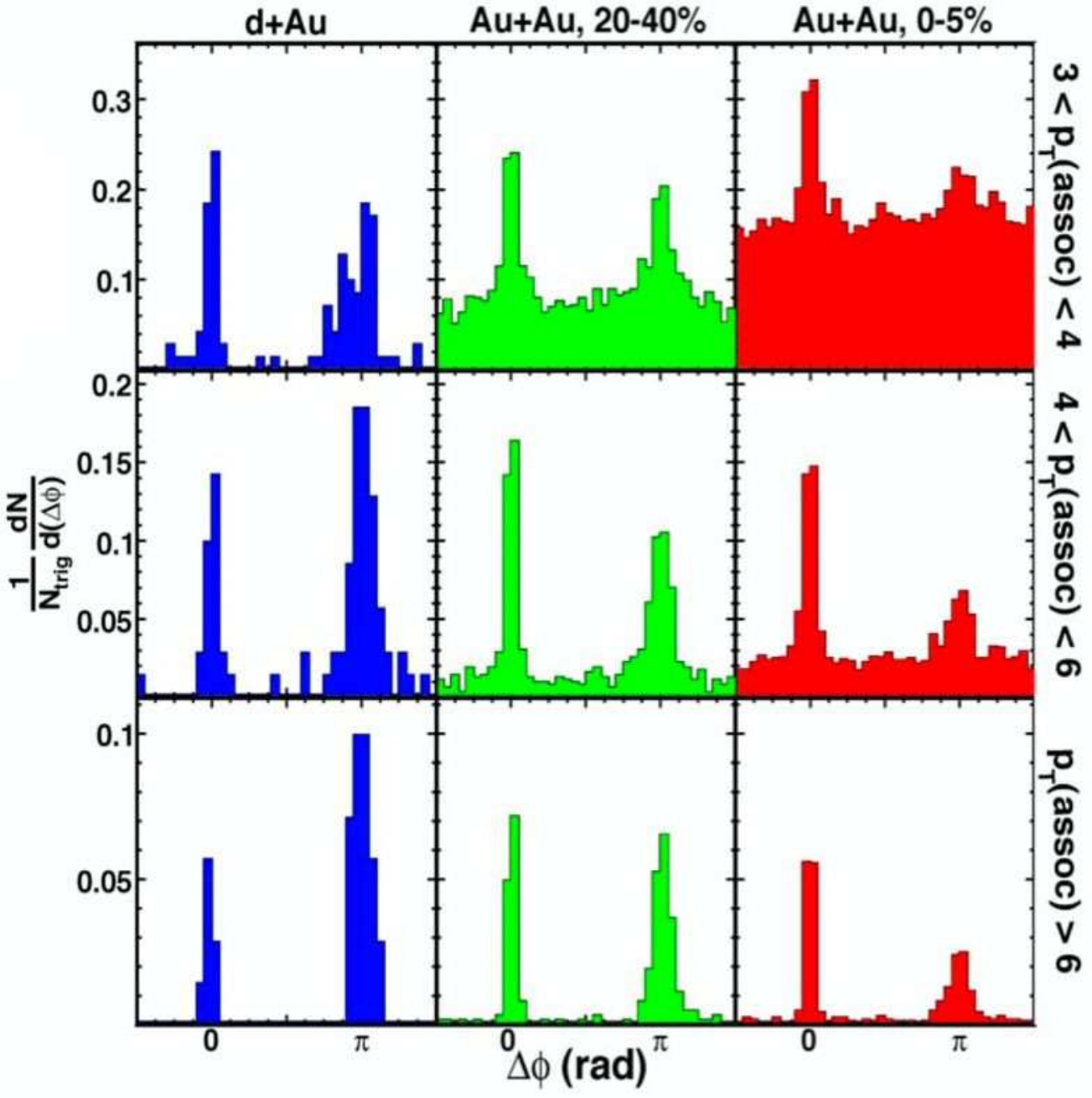}\hspace{2pc}
\raisebox{3pc}{\begin{minipage}[b] {0.4\linewidth}        
\caption[] {Di-hadron azimuthal correlations for charged hadron triggers with $8<p_{T_t}<15$ GeV/c and several values of $p_T^{\rm assoc}$ in minimum bias d+Au and 20\%-40\% (0-5\%) centrality Au+Au collisions at $\sqsn=200$ GeV~\cite{Magestro}}
      \label{fig:Magestrof1}\end{minipage}}
   \end{figure}

However, there is no direct evidence as yet for broadening of di-hadron or di-jet correlations from the effect of $\hat{q}$. The best measurement so far is an early measurement from the STAR collaboration~\cite{Magestro} (Fig.~\ref{fig:Magestrof1}) of di-hadron azimuthal correlations which shows that for a trigger particle with transverse momentum $8<p_{T_t}<15$ GeV/c, the conditional yield of associated charged hadrons with $p_T^{\rm assoc}$ in the away-side peak can be fit to a Gaussian with a width of $\sigma_{\Delta\phi}=0.24\pm 0.07$ for d+Au and $0.20\pm0.02$ ($0.22\pm 0.02$) for 20\%-40\% (0-5\%) centrality Au+Au collisions. 
      %\vspace*{-0.05in}
The result is not statistically significant for a difference in $\sigma_{\Delta\phi}$ for d+Au and Au+Au. 

There are many other such unresolved issues in RHI physics. In the following sections, I review some of the latest results at RHIC starting with some of the original results of hard-scattering and the discovery of ``Jet Quenching'' to place the latest results in context.

\section{Jet Quenching in High \pT Physics at RHIC}
The use of hard-scattering at RHIC as an in-situ probe of the medium produced in A+A collisions by the effect of the medium on outgoing hard-scattered partons has led to many important discoveries (and puzzles). The effect of the medium on outgoing hard-scattered partons produced  \centerline{\mbox{\includegraphics[width=0.75\textwidth]{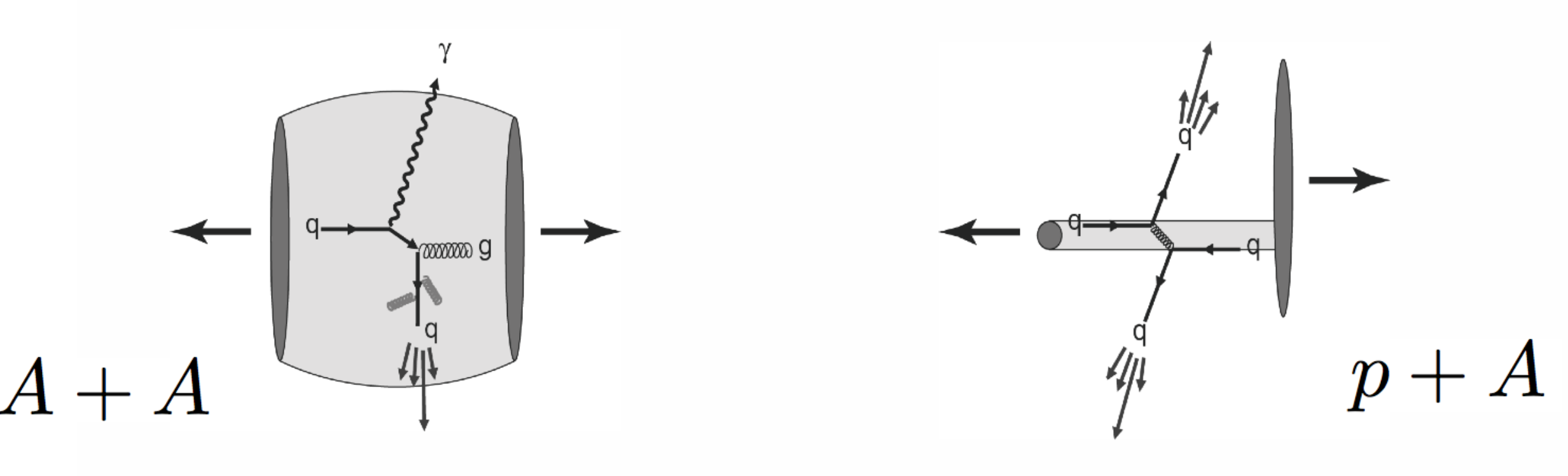} }} 
\mbox{in the  initial}  A+A collision is determined by comparison to measurements in p+A (or d+A) collisions, where no (or negligible) medium is produced.

\begin{figure}[!h] 
      \centering
       \includegraphics[width=0.35\linewidth]{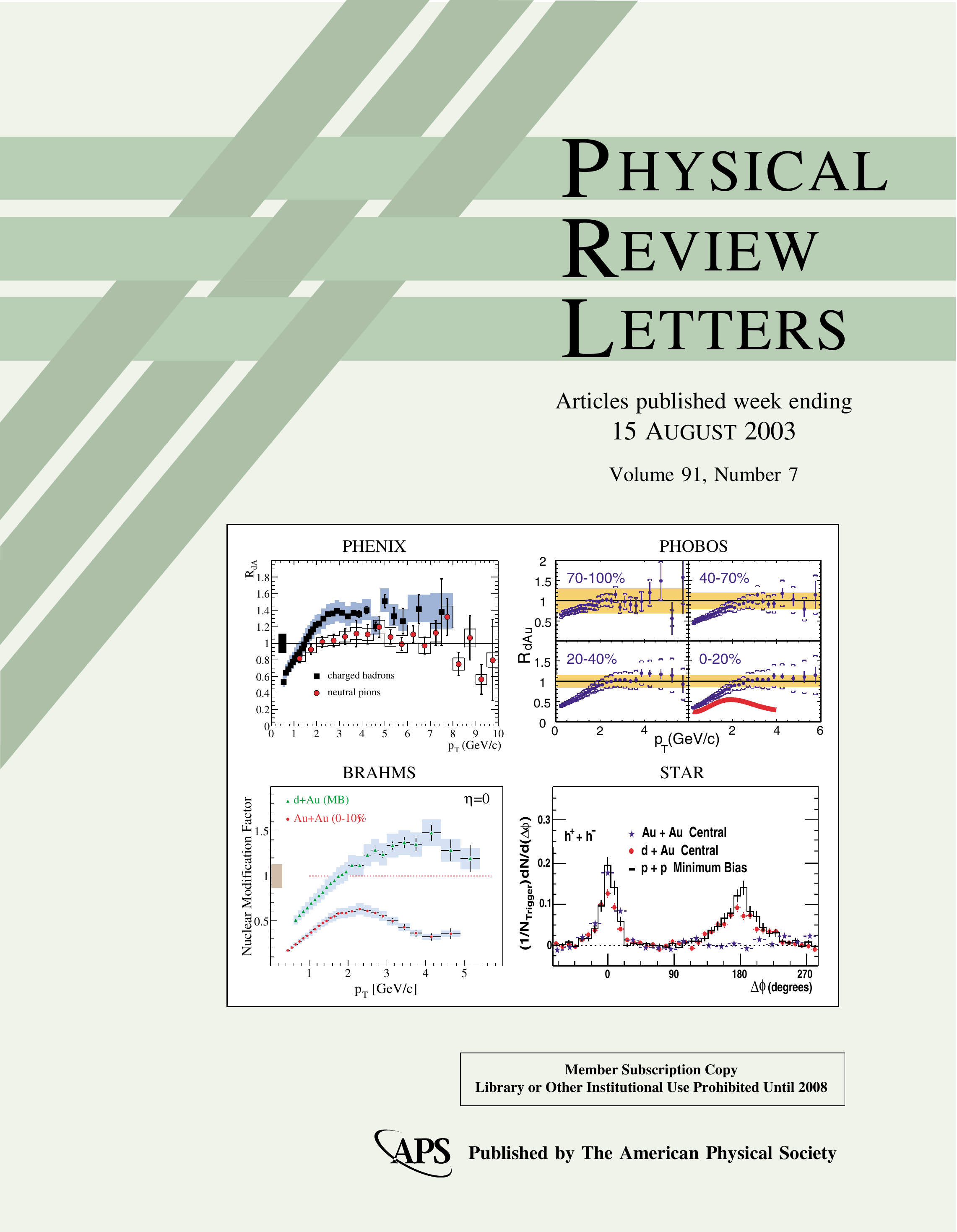}\hspace{1pc}
\raisebox{0.6pc}{\includegraphics[width=0.47\linewidth]{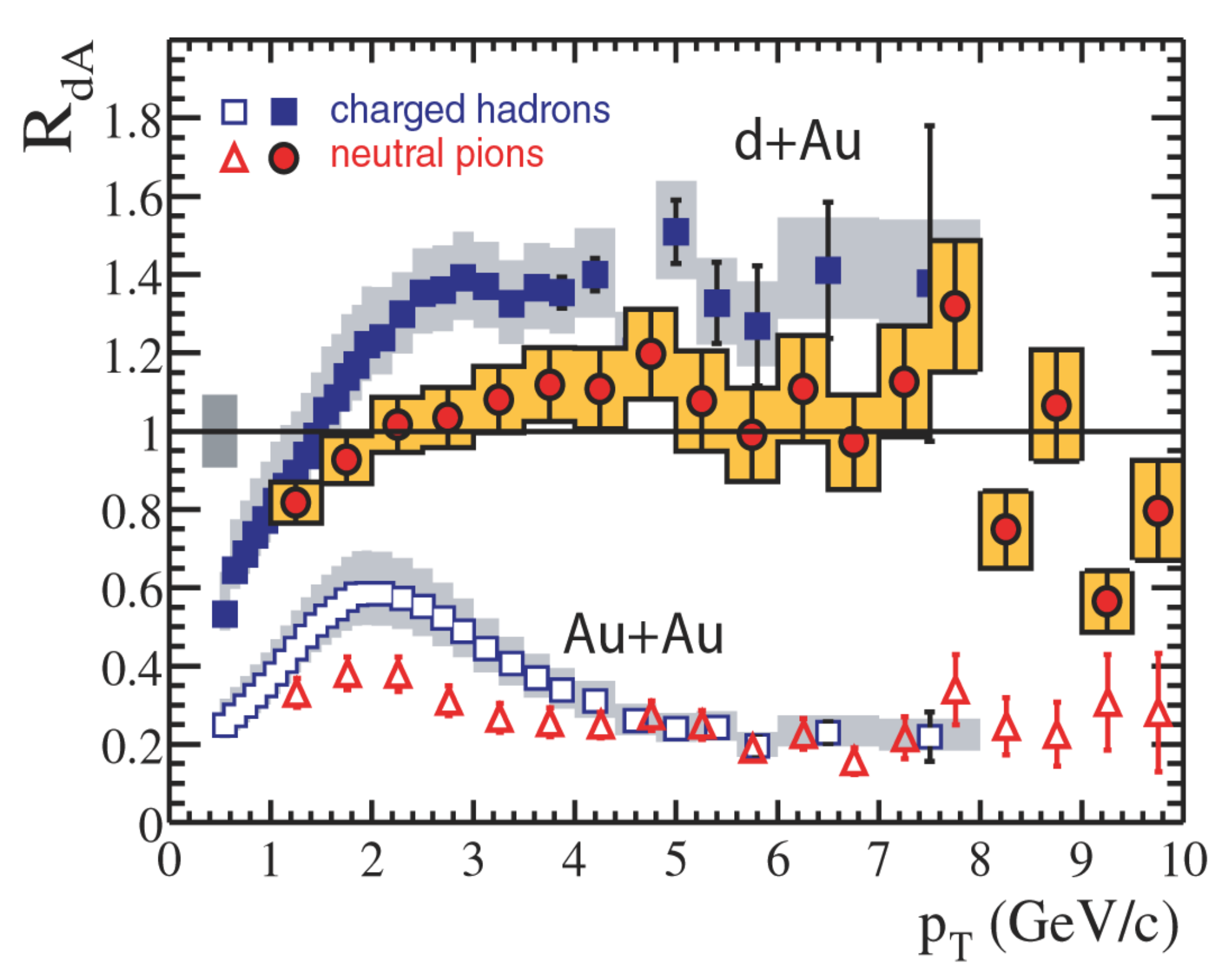}}%%{ij23b}
\caption[] {a) (left) Cover of Phys. Rev. Letters of 13 August 2003 displaying the results of all four RHIC experiments which showed no suppression in d+Au collisions. b) (right) PHENIX results from that volume~\cite{PXPRL91RAA} for d+Au and Au+Au collisions at \sqsn=200 GeV.}
      \label{fig:PRLandPX2003}
   \end{figure}

The discovery at RHIC~\cite{ppg003} that hadrons with $\pT>3$ GeV/c are suppressed in central Au+Au collisions by roughly a factor of 5 compared to point-like (binary) scaling from p-p collisions is arguably {\em the} major discovery in RHI physics. In order to verify that the suppression was due to the medium produced in Au+Au collisions and not an effect in the cold matter of an individual nucleus, measurements in d+Au collisions were performed in 2003 which were so definitive that all four experiments at RHIC had their results displayed on the front cover of Physical Review Letters (Fig.~\ref{fig:PRLandPX2003}a). 
      %\vspace*{-0.05in}
Figure~\ref{fig:PRLandPX2003}b shows the Nuclear Modification factor $\raa$ for non-identified charged hadrons ($h^{\pm}$) and $\pi^0$-mesons measured by PHENIX~\cite{PXPRL91RAA} in d+Au and Au+Au collisions at \sqsn=200 GeV where $\raa(p_T)$ is the ratio of the yield of particles ($h$) for a given centrality Au+Au collision  to the point-like-scaled p-p cross section,
   \begin{equation}
  \raa(p_T)=[{{d^2N^{h}_{AA}/dp_T dy N_{AA}}]/ [{\mean{T_{AA}} d^2\sigma^{h}_{pp}/dp_T dy}}] \quad , 
  \label{eq:RAA}
  \end{equation}
where $\mean{T_{AA}}$ is the overlap integral of the nuclear thickness functions for that centrality. Two things are important to note about Fig.~\ref{fig:PRLandPX2003}b: i) both the $h^{\pm}$ and $\pi^0$ are suppressed in Au+Au central collisions but not suppressed in d+Au collisions; 
ii) the $h^{\pm}$ and $\pi^0$ exhibit different values of $\raa$ in both d+Au and Au+Au collisions. This showed that 
supppression was a medium effect and that particle identification is crucial in these measurements because different particles behave differently. 

\begin{figure}[!t] 
      \centering
             \includegraphics[width=0.42\textwidth]{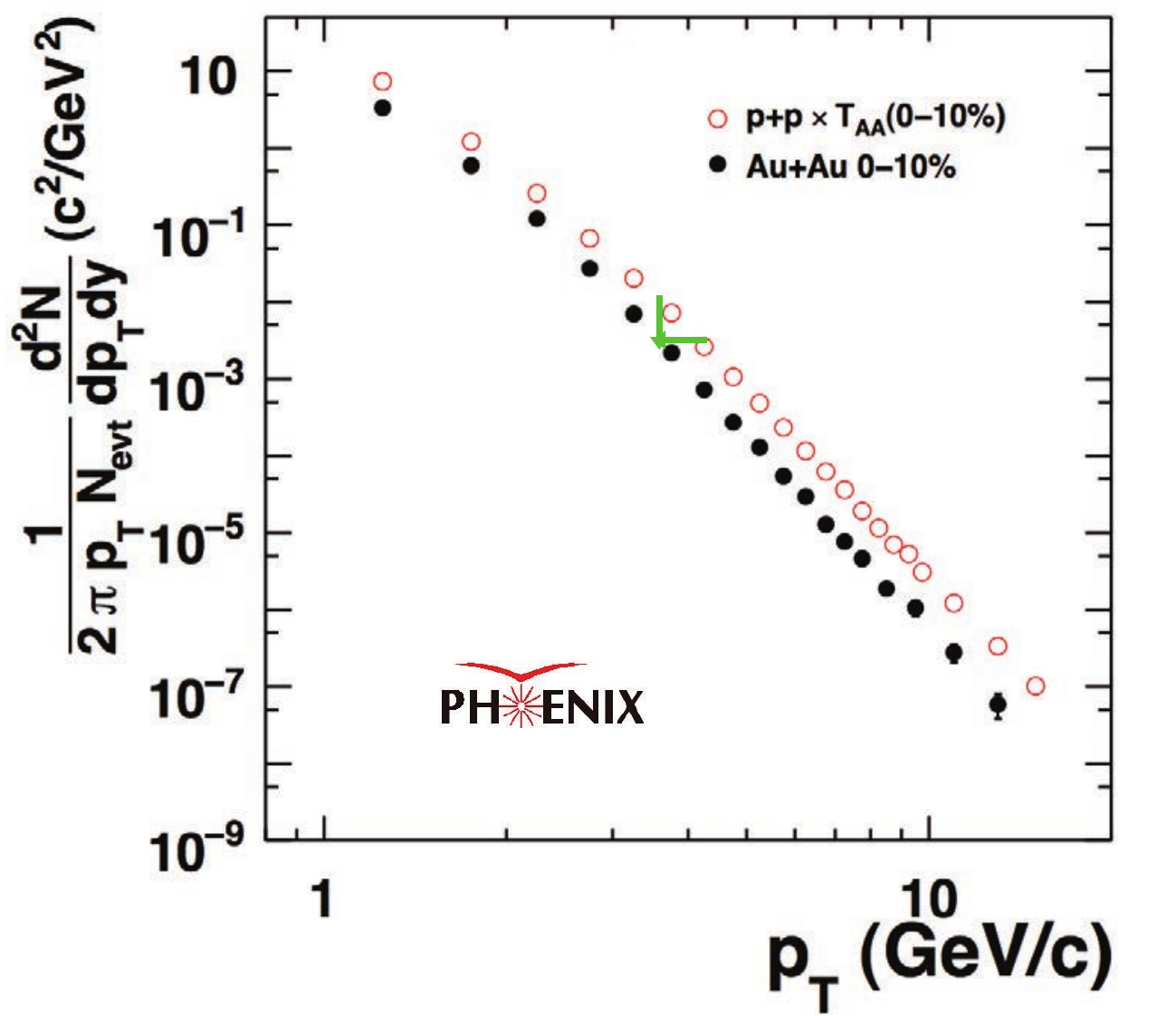}\hspace{1pc}
       \includegraphics[width=0.52\textwidth]{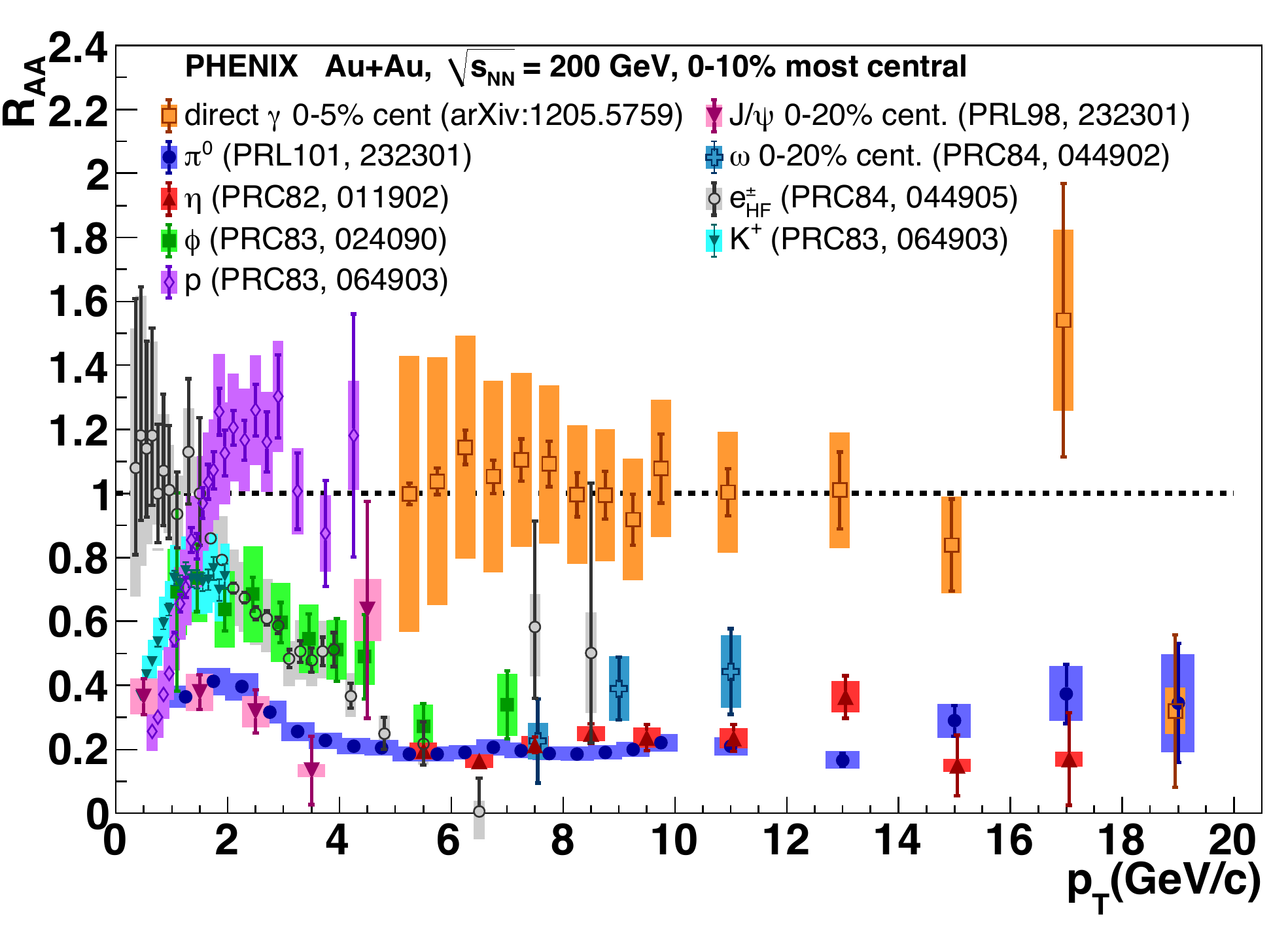}
\caption[] {a) (left) Log-log plot of numerator and denominator of Eq.~\ref{eq:RAA} as a function of \pT for $\pi^0$ at \sqsn=200 GeV~\cite{PXpi0PRC76}  b) PHENIX measurements of \raa(\pT) of many identified particles in central Au+Au collisions at \sqsn=200 GeV, with citations indicated.}\vspace*{-1pc}
      \label{fig:allPXpid}
   \end{figure}

The method of calculating the suppression of high $p_T$ particles in Au+Au collisions at RHIC is shown in Fig.~\ref{fig:allPXpid}a for $\pi^0$~\cite{PXpi0PRC76}. The hard-scattering in p-p collisions is indicated by the power law behavior $p_T^{-n}$ for the invariant cross section, $E d^3\sigma/dp^3$, with $n=8.10\pm 0.05$ for $p_T > 3$ GeV/c at $\sqrt{s_{NN}}=200$ GeV.  The Au+Au data at a given $p_T$ can be characterized either as shifted lower in $p_T$ by $\delta p_T$ from the point-like scaled p-p data at $p'_T=p_T+\delta p_T$, or reduced in magnitude at the same $p_T$, i.e. suppressed. In Fig.~\ref{fig:allPXpid}b, the suppression of the many identified particles measured by PHENIX at RHIC over many years is presented as the Nuclear Modification Factor, \raa . 
These results show an interesting pattern. 
All particles are suppressed for $\pT\geq 2$ GeV/c, even the electrons ($e^\pm_{\rm HF}$) from $c$ and $b$ quark decay 
(but not the direct-$\gamma$). There is one notable exception---the protons are enhanced. This is called the baryon anomaly~\cite{ppg015,ppg023}.
The non-suppression of the direct-$\gamma$ in Au+Au collisions while the $\pi^0$, $\eta$ and all the other mesons in Fig.~\ref{fig:allPXpid} are suppressed shows that the suppression is an effect of the medium, the \QGP, presumably due to the loss of energy by their parent quark or gluon with exposed color charge.  

      \begin{figure}[!b] 
      \centering
       \includegraphics[width=0.55\linewidth]{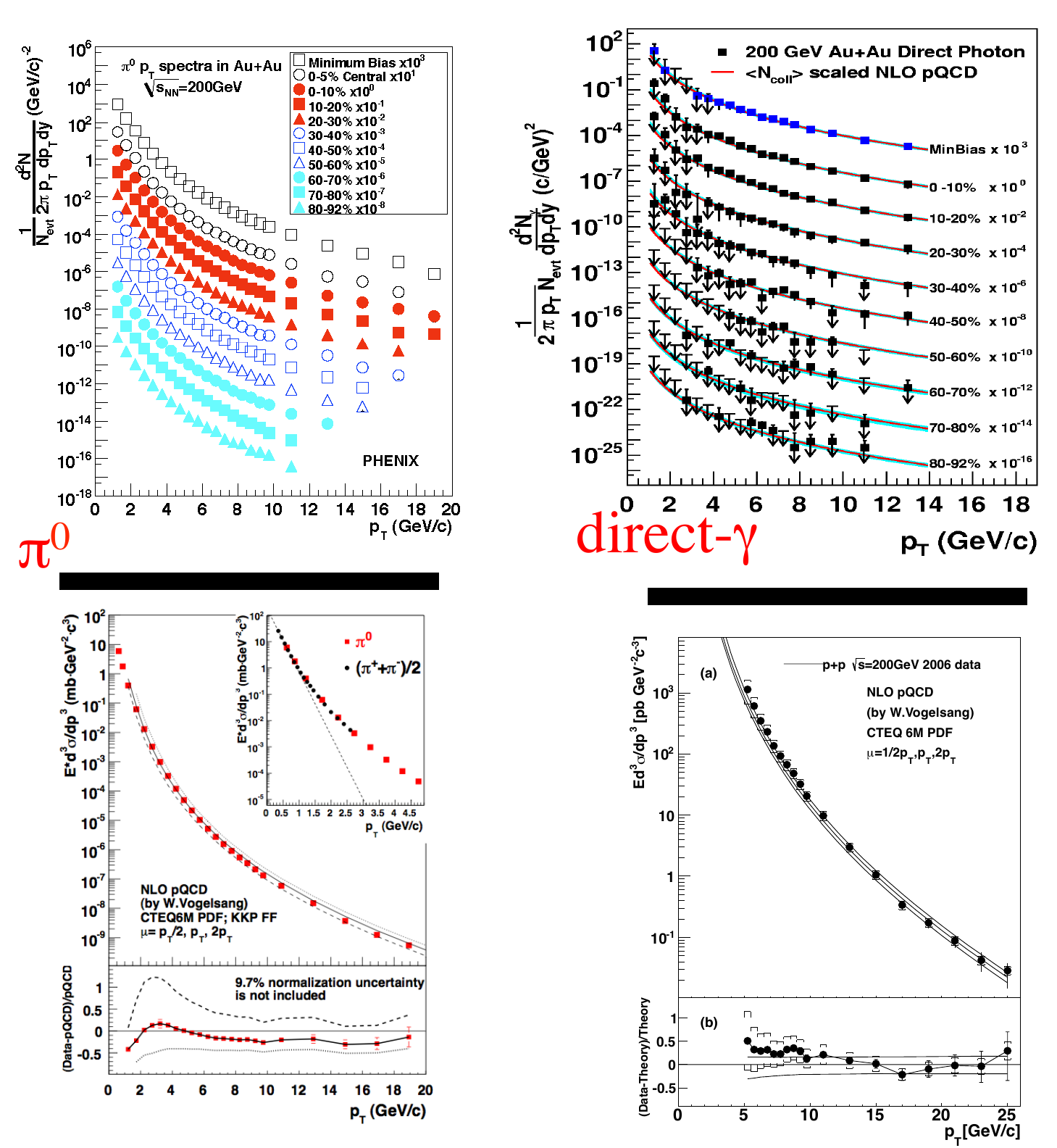} \hspace{1pc}
\raisebox{3.3pc}{\begin{minipage}[b]{0.40\linewidth} \includegraphics[width=\linewidth]{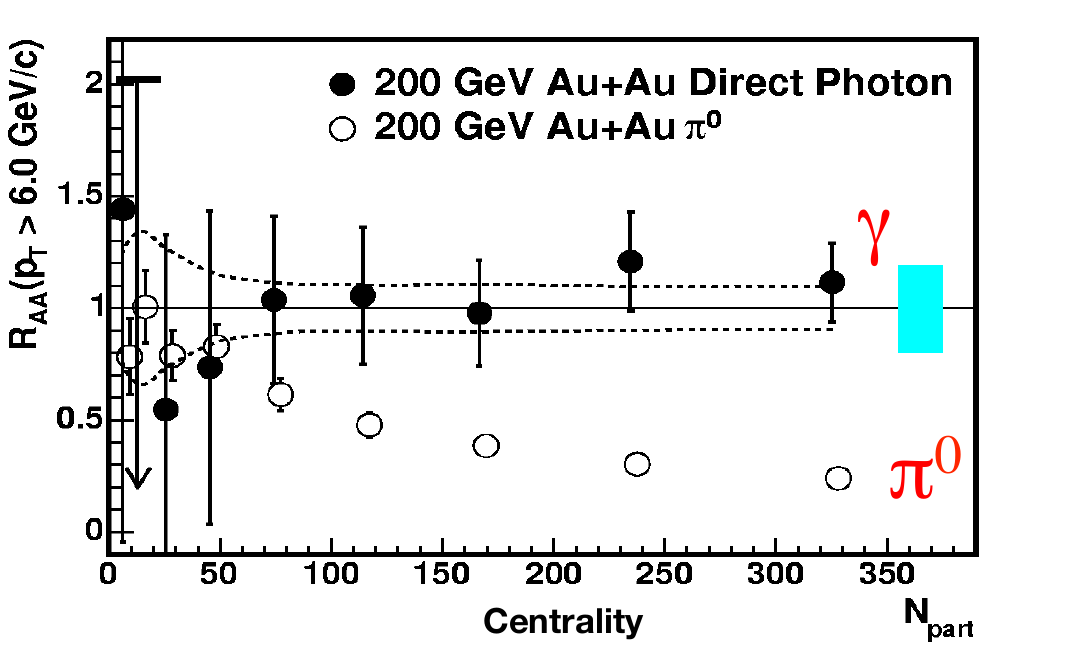} 
\caption[] {a) (left) PHENIX measurements of $\pi^0$ in Au+Au\cite{PXPRL101}/p-p\cite{PXpi0PRD76} and direct-$\gamma$ in Au+Au\cite{PXdirgPRL94}/p-p\cite{PXdirgppPRD86}. b) \raa({\pT}) for $\pi^0$ and \mbox{direct-$\gamma$} as a function of centrality~\cite{PXdirgPRL94} from the measurements in (a).}
      \label{fig:4msmts} \end{minipage}}
   \end{figure}
Direct-$\gamma$ production at large \pT is one of most beautiful hard-scattering processes in \QCD~\cite{FritzschMinkowski} as well as the best probe of the \QGP\ because the $\gamma$ ray participates directly in the predominant constituent reaction $g+q\rightarrow \gamma+q$ and emerges uscathed by the medium so that its momentum can be measured precisely. Also, since the scattered quark has equal and opposite \pT to the direct-$\gamma$, the \pT of the outgoing quark is known precisely at the production point. The simplicity and clarity of the constituent reaction stands in marked contrast to the difficulty of the experiment due to the overwhelming background of $\gamma$-rays from the $\gamma +\gamma$ decay of $\pi^0$ and $\eta$ mesons. However, with a properly designed experiment one can make four measurements, the $\pi^0$ and \mbox{direct-$\gamma$} \pT spectra in both p-p and A+A collisions which span many orders of magnitude (Fig.~\ref{fig:4msmts}a), to obtain a precise comparison of \raa for $\pi^0$ and direct-$\gamma$ (Fig.~\ref{fig:4msmts}b)~\cite{PXdirgPRL94}. The $\pi^0$ are more suppressed with increasing centrality while the direct-$\gamma$, with constant $\raa\approx1$, are unaffected by the medium.   

\begin{figure}[!h] 
      \centering
       \includegraphics[width=0.48\linewidth]{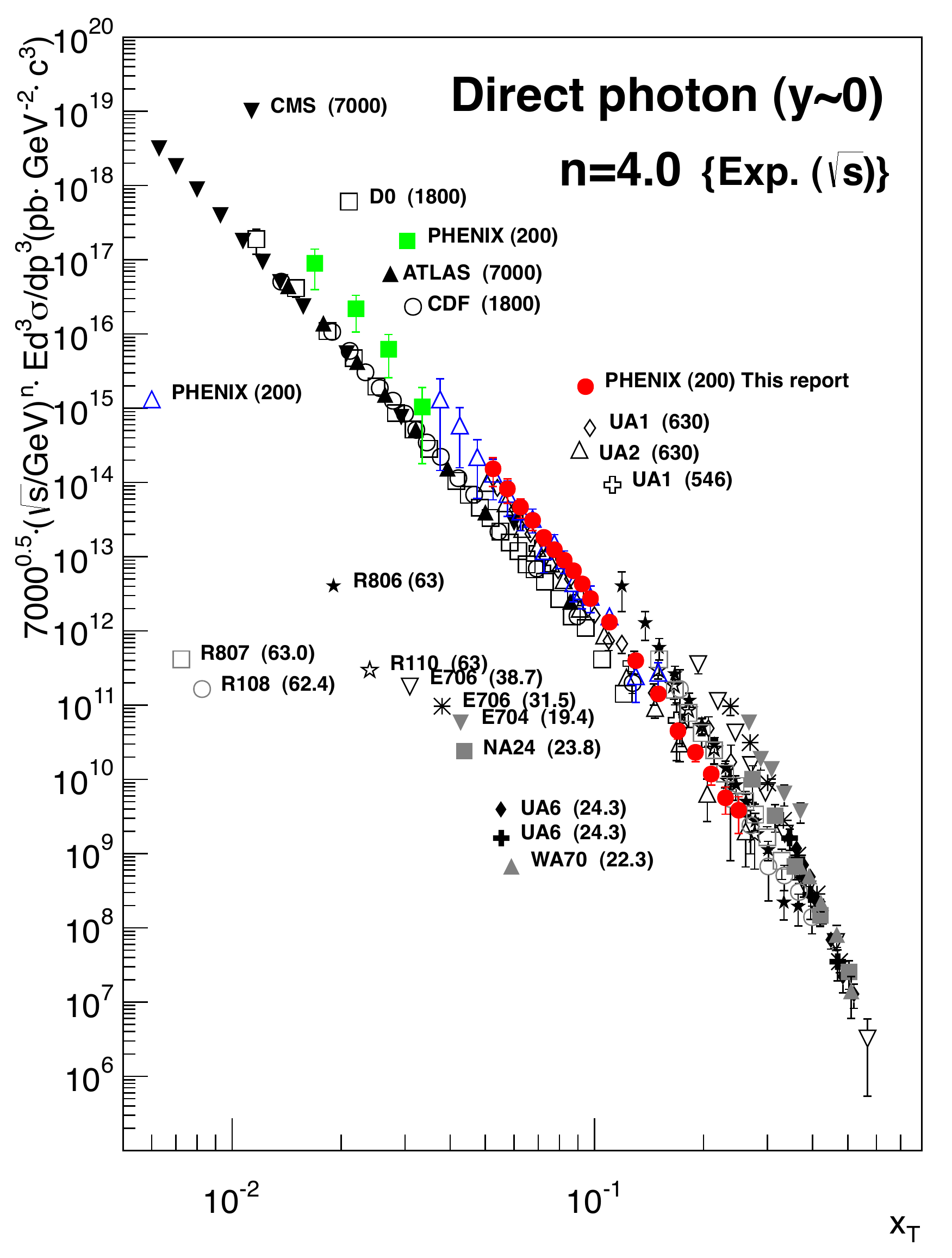}\hspace{1pc}
\raisebox{0.0pc}{\includegraphics[width=0.48\linewidth]{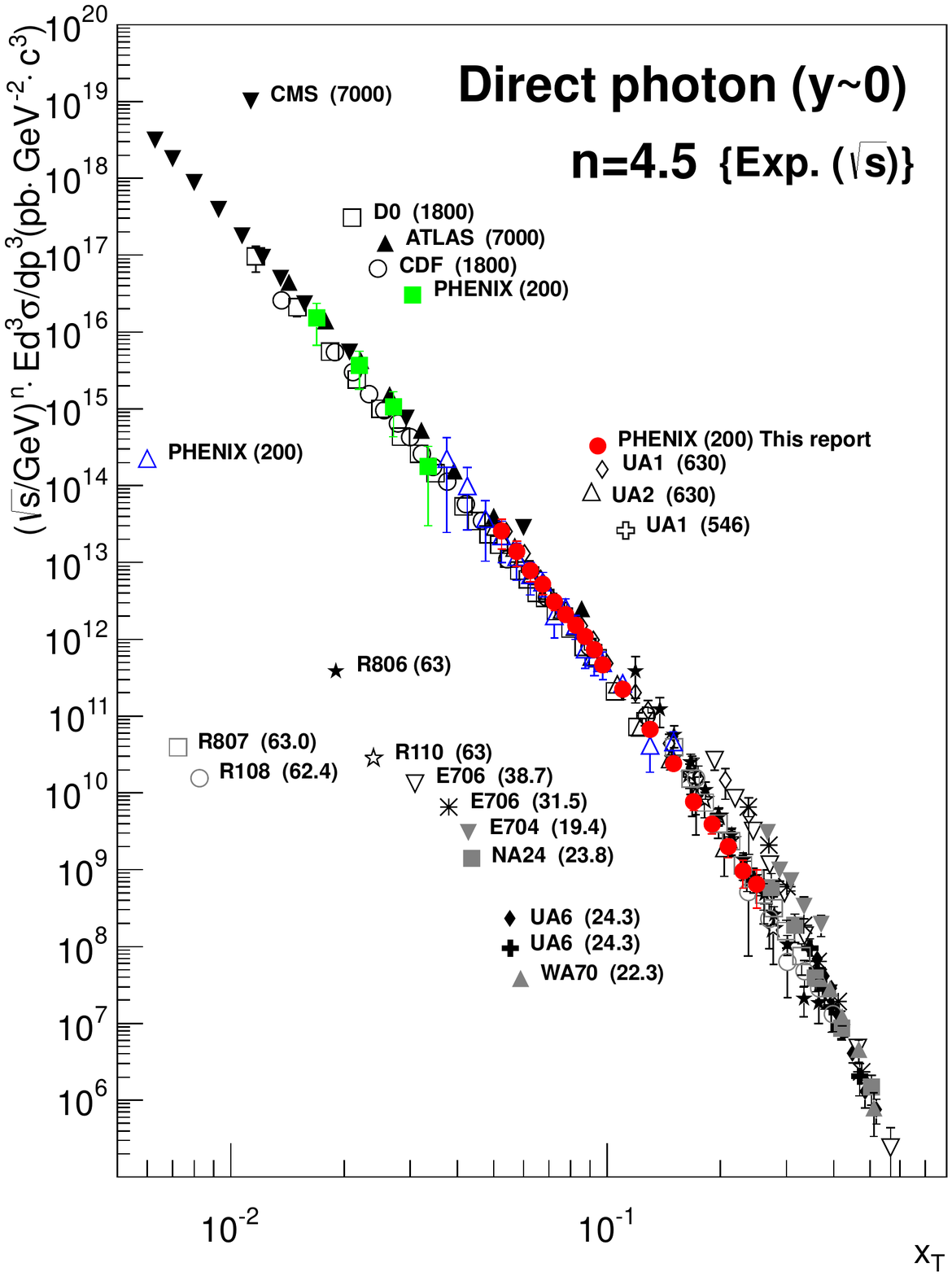}}%%{ij23b}
\caption[] {Direct-$\gamma$ measurements plotted as $\sqrt{s}^{\, n_{\rm eff}} \times Ed^3\sigma/dp^3 (x_T)$: a) (left) with $n_{\rm eff}=4.0$ b) (right) with $n_{\rm eff}=4.5$; where $x_T=2\pT/\sqs$~\cite{PXdirgppPRD86}. The legend gives the experiment and $\sqrt{s}$. }
      \label{fig:alldirgxT}
   \end{figure}

The dominance of the single constituent subprocess $g+q\rightarrow \gamma +q$ for direct-$\gamma$ production allows an illustration of a fundamental property of \QCD\ using the recent PHENIX measurements in p-p at \sqs=200 GeV~\cite{PXdirgppPRD86} combined with a collection of the world's measuements of direct-$\gamma$ production in p-p and $\bar{\rm p}$+p collisions. Figure~\ref{fig:alldirgxT}a shows an $x_T$ scaling plot of all the measurements using the naive parton model value of $n_{\rm eff}=4.0$~\cite{BBK,BBGPLB42}, which shows significant deviations compared to the beautiful $x_T$ scaling (Fig.~\ref{fig:alldirgxT}b) with index $n_{\rm eff}=4.5$.  The difference of Figs.~\ref{fig:alldirgxT}a and b illustrates the non-scaling of \QCD\ due to the running coupling constant $\alpha_s(Q^2)$ and the evolution of the parton distribution functions~\cite{CahalanPRD11}. 

      %\vspace*{-0.05in}

For those who find the discussion in this section too sketchy, Jan Rak and I have published a book this year, with all this kind of information, which gives a full survey of ``High \pT Physics in the Heavy Ion Era'' over more than a century, from Rutherford (1911) to 2012, including many of the Pb+Pb results from the LHC~\cite{RTbook}.

\section{New Results since Utrecht-2011}
\begin{figure}[!t]
   \begin{center}
\includegraphics[width=0.49\textwidth,height=0.38\textwidth]{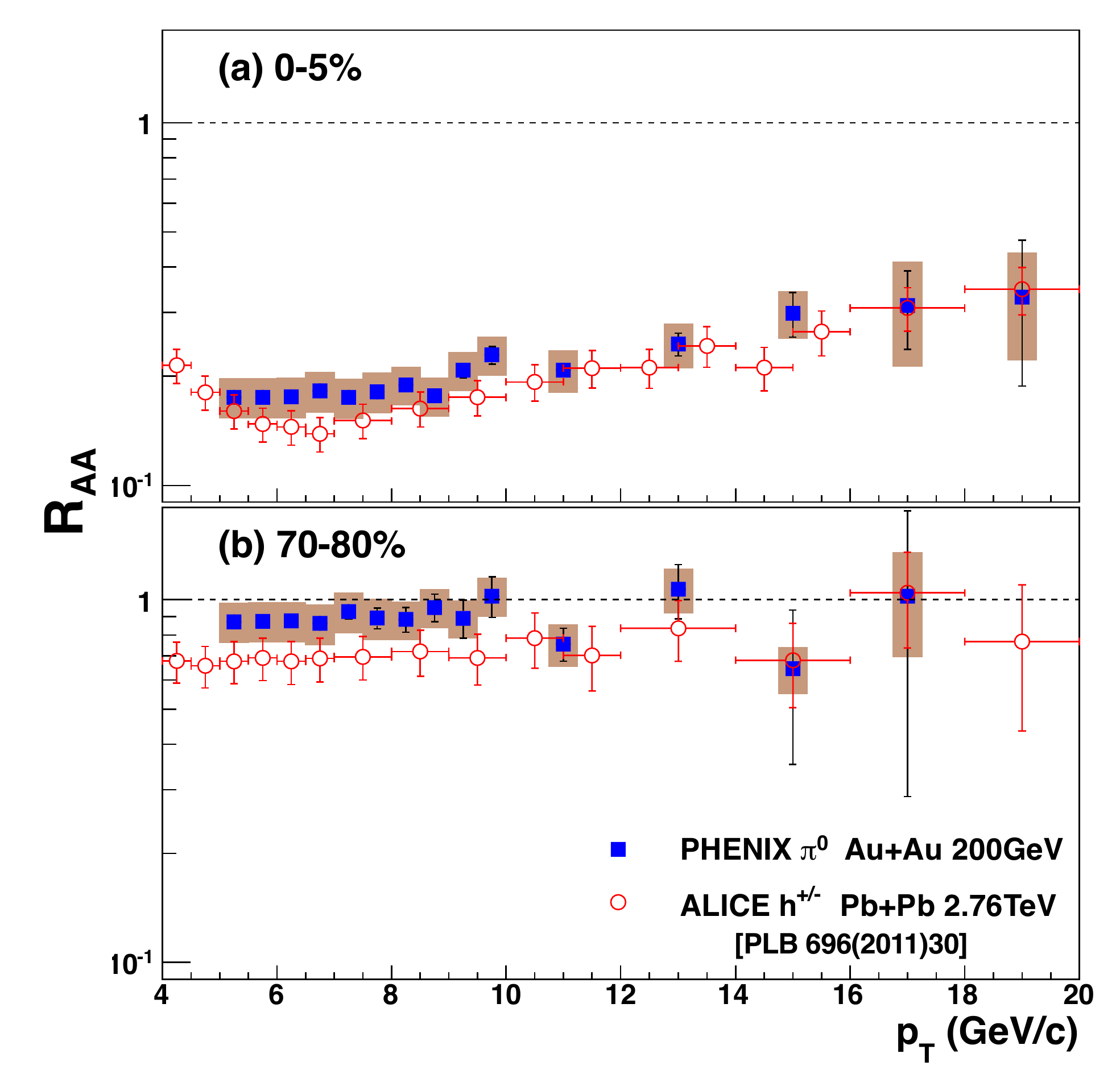}
\includegraphics[width=0.49\textwidth]{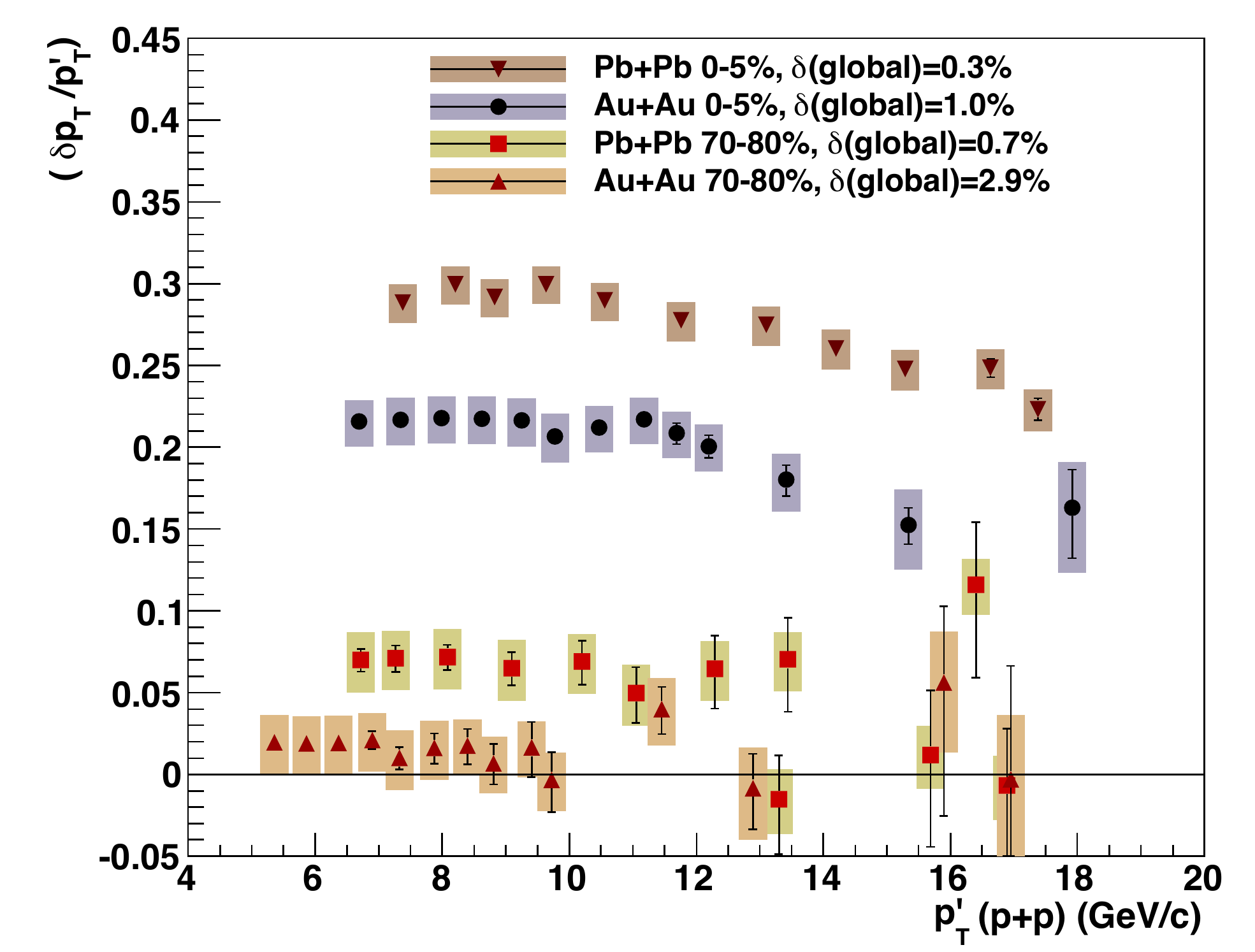}
\end{center}\vspace*{-1.7pc}
\caption[]{a) (left) $R_{AA}$ of $\pi^0$ in $\sqrt{s_{NN}}=200$ GeV central (0-5\%) and peripheral (70-80\%) Au+Au collisions~\cite{ppg133} at RHIC compared to non-identified charged hadron (${\rm h}^{\pm}$) $R_{AA}$ in $\sqrt{s_{NN}}=2.76$ TeV Pb+Pb collisions at LHC. b) (right) Fractional shift of $p_T$ spectrum $\delta p_T/p'_T$ vs. $p'_T$ (p-p) calculated by PHENIX~\cite{ppg133} for RHIC and LHC. 
\label{fig:ppg133}}\vspace*{-0.8pc}
\end{figure}

\begin{figure}[!b]
   \begin{center}
\includegraphics[width=0.99\textwidth]{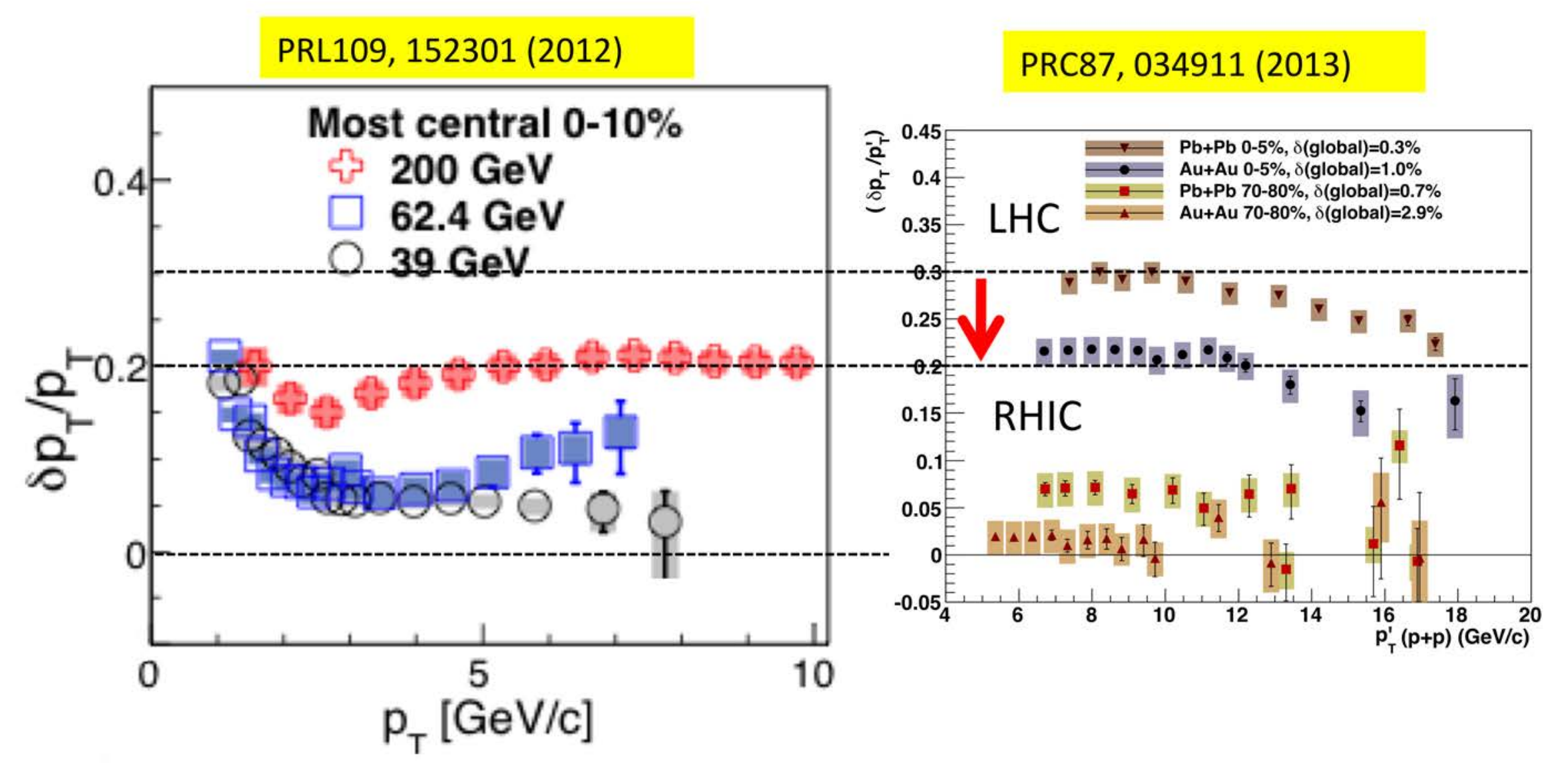}
\end{center}\vspace*{-1.7pc}
\caption[]{Fractional shift of $p_T$ spectrum $\delta p_T/p'_T$ in central A+A collisions from $\sqrt{s_{NN}}=39$~GeV to 2.76 TeV 
\label{fig:shiftallRHICLHC}}\vspace*{-1.8pc}
\end{figure}

  At the last meeting in this series that I attended, in Utrecht, the first results of the ALICE measurements of suppression of high $p_T$ charged hadrons $h^{\pm}$ at \sqsn=2.76 TeV at LHC were presented~\cite{ALICE2010}. Despite more than a factor of 20 higher c.m. energy, the $R_{AA}$ measurements by ALICE at LHC (Fig.~\ref{fig:ppg133}a)~\cite{ppg133} appear to be nearly identical to those from PHENIX at RHIC for $5\leq p_T\leq 20$ GeV/c, except that for the LHC data, with better statistics, the upward trend of $R_{AA}(p_T)$ over the whole interval is significant (although the more recent PHENIX data in Fig.~\ref{fig:ppg133}a~\cite{ppg133} show a small but significant upward trend).  
However, what is most important to realize is that  
the exponent of the power-law at LHC ($n\approx 6$) is flatter than at RHIC ($n\approx 8$), so that to get the same value for $R_{AA}$ at LHC as at RHIC, a $\sim 40$\% larger shift $\delta p_T/p'_T$ in the spectrum from p-p to A+A is required. This led us more recently to compare the data in terms of $\delta p_T/p'_T$ (Fig.~\ref{fig:ppg133}b) which clearly shows the larger shift for central collisions at LHC. The larger shift in the \pT\ spectrum likely indicates $\sim 40$\% larger fractional energy loss at LHC than at RHIC in this $p_T$ range due to the probably hotter and denser medium. These measurements can be combined with the previous measurements at RHIC for $\sqsn=39$ and 62.4 GeV~\cite{ppg138} (Fig.~\ref{fig:shiftallRHICLHC}) to reveal a systematic increase of $\delta p_T/p'_T$ in central A+A collisions at $p'_T=7$ GeV/c, going from 5\% to 30\% over the c.m. energy range $\sqsn=39$~GeV to 2.76 TeV.   
  
\begin{figure}[!h]
\includegraphics[height=0.22\textheight]{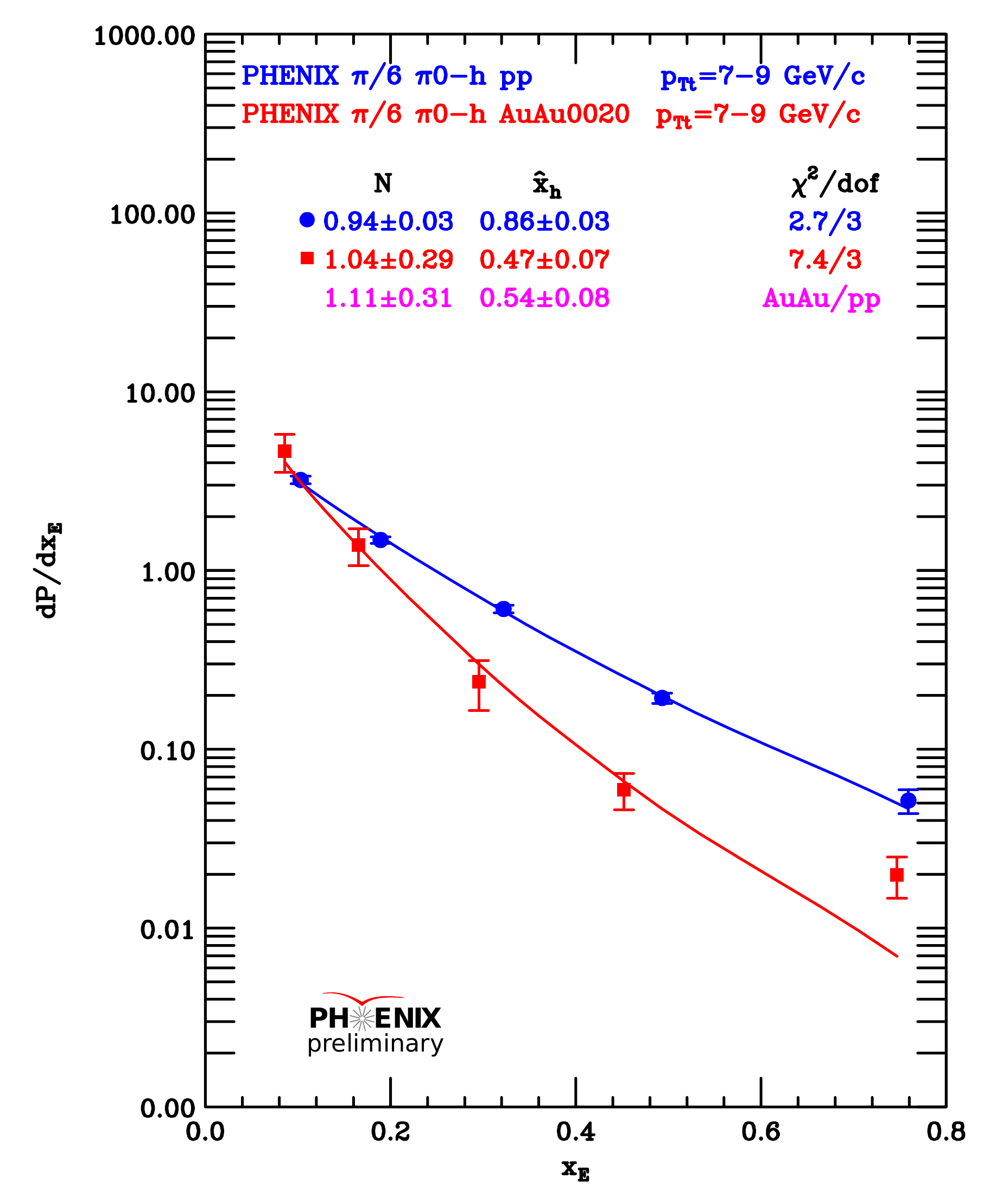}
\includegraphics[height=0.22\textheight]{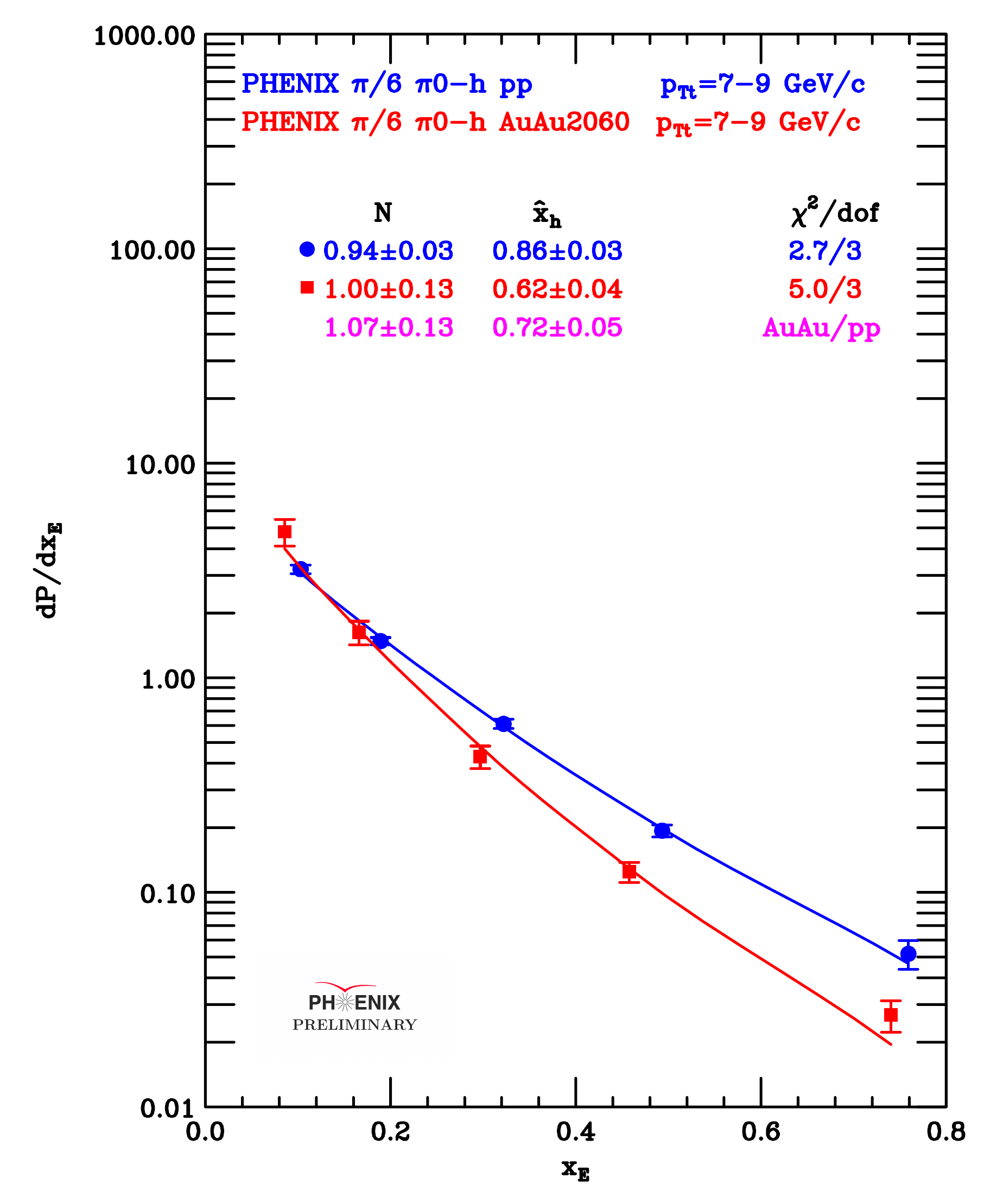}
\includegraphics[height=0.22\textheight]{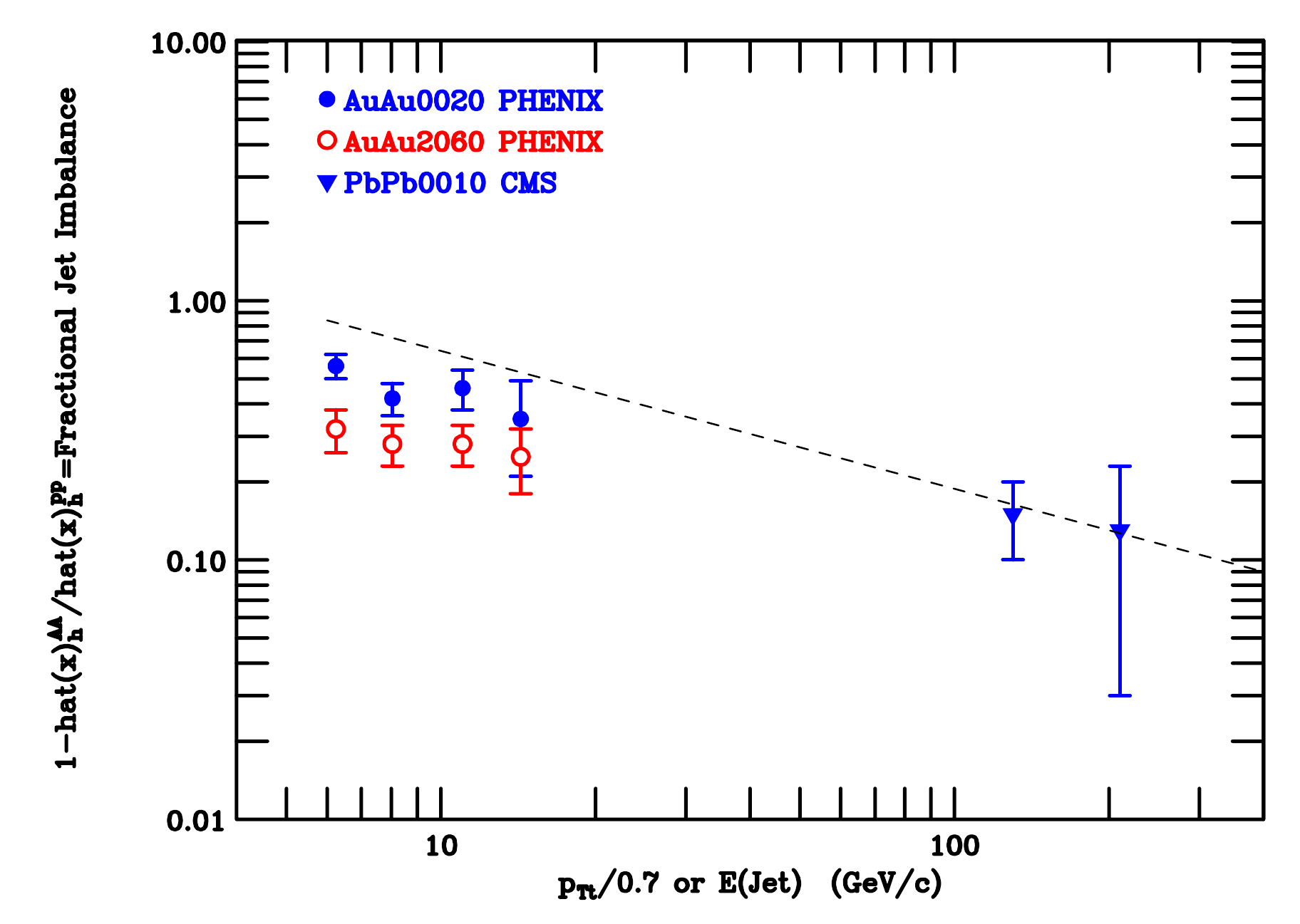}%\vspace*{-1pc}
\caption[]{(left) $x_E$ distributions at RHIC~\cite{MJT-Utrecht} from p-p (blue circles) and AuAu (red squares) collisions for $p_{T_t}=7-9$ GeV/c, together with fits to Eq.~\ref{eq:condxeN2} with parameters indicated: p-p (solid blue line) $N^{pp}=0.94\pm0.03$, $\hat{x}_h^{pp}=0.86\pm0.03$. For AuAu (solid red line), the ratios of the fitted parameters for AuAu/pp are also given:   a) 00-20\% centrality, $N^{AA}=1.04\pm0.29$, $\hat{x}_h^{AA}=0.47\pm0.07$, $\hat{x}_h^{AA}/\hat{x}_h^{pp}=0.54\pm0.08$;  b) 20--60\% centrality $N^{AA}=1.00\pm0.13$, $\hat{x}_h^{AA}=0.62\pm0.04$, $\hat{x}_h^{AA}/\hat{x}_h^{pp}=0.72\pm0.05$. c) (right) Fractional jet imbalance~\cite{MJT-Utrecht}, $1-\hat{x}_h^{AA}/\hat{x}_h^{pp}$, for the RHIC data from (a) and (b), and CMS data~\cite{CMSPRC84} as estimated by MJT~\cite{MJT-Utrecht}. The dashed line is an estimate~\cite{MJT-Utrecht} of the Fractional Jet Imbalance vs. E(Jet) at \sqsn=2.76 TeV. }
\label{fig:AuAupp79}
\end{figure}

\pagebreak  
     At Utrecht, I also discussed another unexpected discovery at RHIC~\cite{ppg029}, that the distribution of particles, with $p_{T_a}$, opposite in azimuth to a trigger particle, e.g. a $\pi^0$ with large $p_{T_t}$, which is itself the fragment of a jet, does not measure the fragmentation function of the jet opposite in azimuth to the trigger, but, instead, measures the ratio of $\hat{p}_{T_a}$, of the away-parton, to $\hat{p}_{T_t}$, of the trigger-parton, and depends only on the same power $n$ as the invariant single particle spectrum:  \vspace*{-0.5pc}
\begin{equation}
{dP / dx_E}|_{p_{T_t}} \approx {N\,(n-1)}/[\,\hat{x}_h 
{(1+ x_E/\hat{x}_h})^{n}] \qquad .\label{eq:condxeN2}\vspace*{-0.5pc}\end{equation}  
This equation gives a simple relationship between the ratio, $x_E\approx p_{T_a}/p_{T_t}$, of the transverse momenta of the away-side particle to the trigger particle, and the ratio of the transverse momenta of the away-jet to the trigger-jet, $\hat{x}_{h}=\hat{p}_{T_a}/\hat{p}_{T_t}$. PHENIX measurements~\cite{PXpi0hPRL104} of the $x_E$ distributions of $\pi^0$-h correlations in p-p and Au+Au collisions at $\sqrt{s_{NN}}=200$ GeV were fit to Eq.~\ref{eq:condxeN2} (Fig.~\ref{fig:AuAupp79}a,b)~\cite{MJT-Utrecht}.
The results for the fitted parameters are shown on the figures. In general the values of $\hat{x}^{pp}_h$ do not equal 1 but vary between $0.8<\hat{x}^{pp}_h<1.0$ due to $k_T$ smearing and the range of $x_E$ covered. In order to take account of the imbalance ($\hat{x}^{pp}_h <1$) observed in the p-p data, the ratio $\hat{x}_h^{AA}/\hat{x}_h^{pp}$ is taken as the measure of the energy of the away jet relative to the trigger jet in A+A compared to p-p collisions. 

\begin{figure}[!h]
\begin{center}
\includegraphics[width=0.72\textwidth]{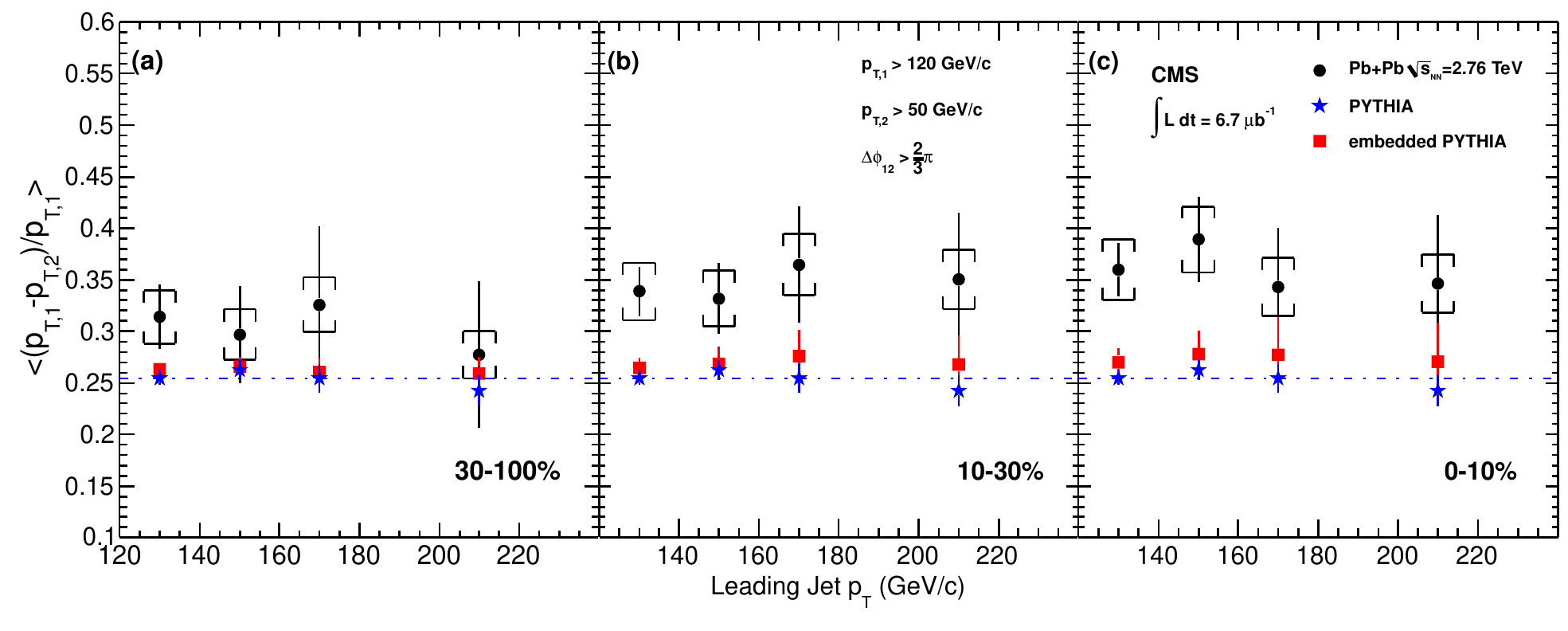}
\end{center}\vspace*{-2pc}
\caption[]{CMS measurement~\cite{CMSPRC84} of $\mean{1-\hat{p}_{t,2}/\hat{p}_{t,1}}$, the fractional jet imbalance, for 3 centralities in Pb+Pb collisions at $\sqrt{s_{NN}}=2.76$ TeV, compared to PYTHIA simulations for p-p collisions.}
\label{fig:CMSAJ2011}
\end{figure}

\begin{figure}[!h]
\begin{center}
\includegraphics[width=0.66\textwidth]{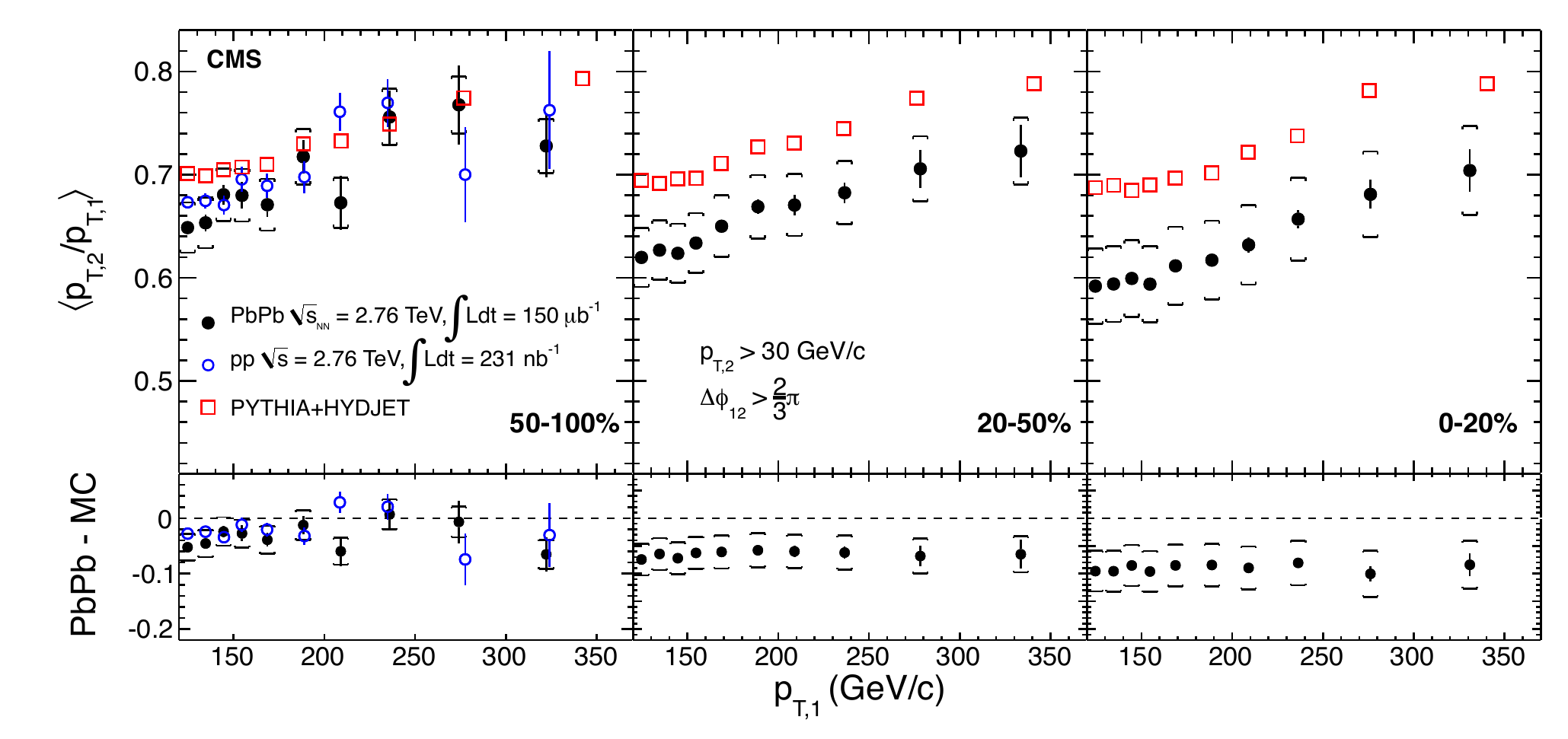}
\raisebox{1pc}{\includegraphics[height=0.24\textwidth]{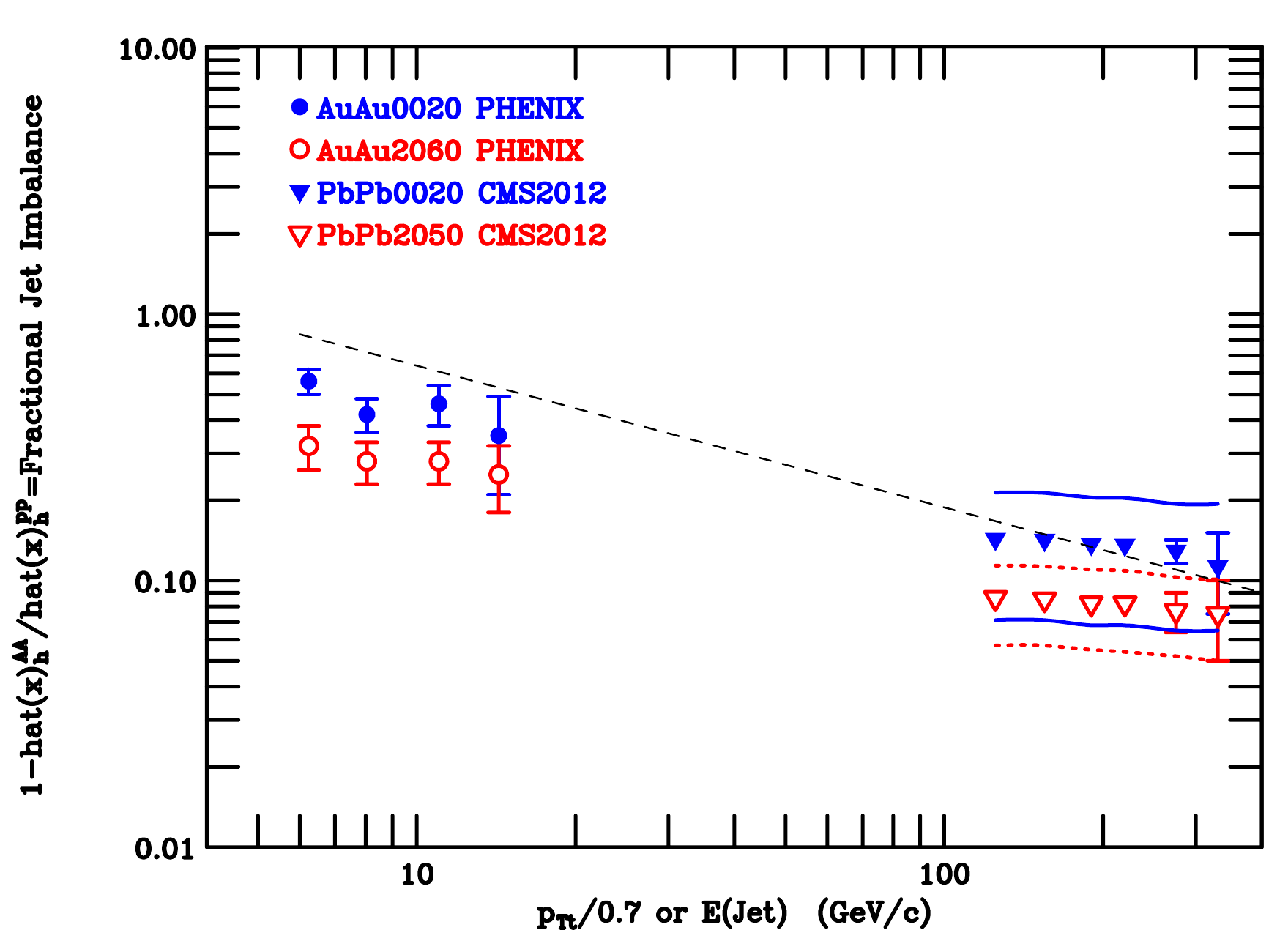}}
\end{center}\vspace*{-2pc}
\caption[]{a) (left 3 panels) CMS~\cite{CMS-dijet-PLB712} measurements of average di-jet transverse momentum ratio,  $\hat{x}_h=\hat{p}_{T,2}/\hat{p}_{T,1}$, as a function of leading jet $\hat{p}_{T,1}$ at $\sqrt{s_{NN}}=2.76$ TeV in p-p collisions and for 3 centralities in Pb+Pb collisions, as well as simulated p-p di-jets embedded in heavy ion events. b) (right) Fractional jet imbalance~\cite{MJT-Utrecht}, $1-\hat{x}_h^{AA}/\hat{x}_h^{pp}$, for the RHIC data from Fig.~\ref{fig:AuAupp79} with CMS measurement from (a). The solid (dotted) lines represent the systematic uncertainty of the CMS 0-20\% (20-50\%) results. The dashed line is the estimate of the Fractional Jet Imbalance vs. E(Jet) at \sqsn=2.76 TeV from Fig.~\ref{fig:AuAupp79}c. }   
\label{fig:CMSdijet2012}
\end{figure}

The fractional jet imbalance was also measured directly with reconstructed di-jets by the CMS collaboration at the LHC in Pb+Pb central collisions at $\sqrt{s_{\rm NN}}=2.76$ TeV (Fig.~\ref{fig:CMSAJ2011})~\cite{CMSPRC84}. At Utrecht, I 
calculated the same ratio $\hat{x}_h^{AA}/\hat{x}_h^{pp}$ for the CMS data to correct for the large effect in p-p collisions~\cite{MJT-Utrecht} and the results were compared to PHENIX as shown in Fig.~\ref{fig:AuAupp79}c. The dashed line shows my estimate of the Fractional Jet Imbalance vs. E(Jet) at \sqsn=2.76 TeV and includes the 40\% increase in $\delta p_T/p'_T$ from Fig.~\ref{fig:ppg133}b.
New results in 2012 by CMS (Fig.~\ref{fig:CMSdijet2012})~\cite{CMS-dijet-PLB712} significantly extend and improve their previous measurement and confirm my correction~\cite{MJT-Utrecht}. 

The large difference in fractional jet imbalance between RHIC and LHC c.m. energies could be due to the difference in jet $\hat{p}_{T_t}$ between RHIC ($\sim 20$ GeV/c) and LHC ($\sim 200$ GeV/c), a difference in the properties of the medium, the difference in $n$ for the different $\sqrt{s_{NN}}$, or a problem with Eq.~\ref{eq:condxeN2} which has not been verified by direct comparison to di-jets. In any case the strong $\hat{p}_T$ dependence of the fractional jet imbalance (apparent energy loss of a parton) also seems to disfavor purely radiative energy-loss in the \QGP~\cite{BDMPSZ} and indicates that the details of energy loss in a \QGP\ remain to be understood.  Future measurements at both RHIC and LHC will need to sort out these issues by extending di-jet and two-particle correlation measurements to overlapping regions of $\hat{p}_T$.   

\begin{figure}[!h] 
      \centering
\raisebox{0pc}{\begin{minipage}[b]{0.45\linewidth} \includegraphics[width=\linewidth]{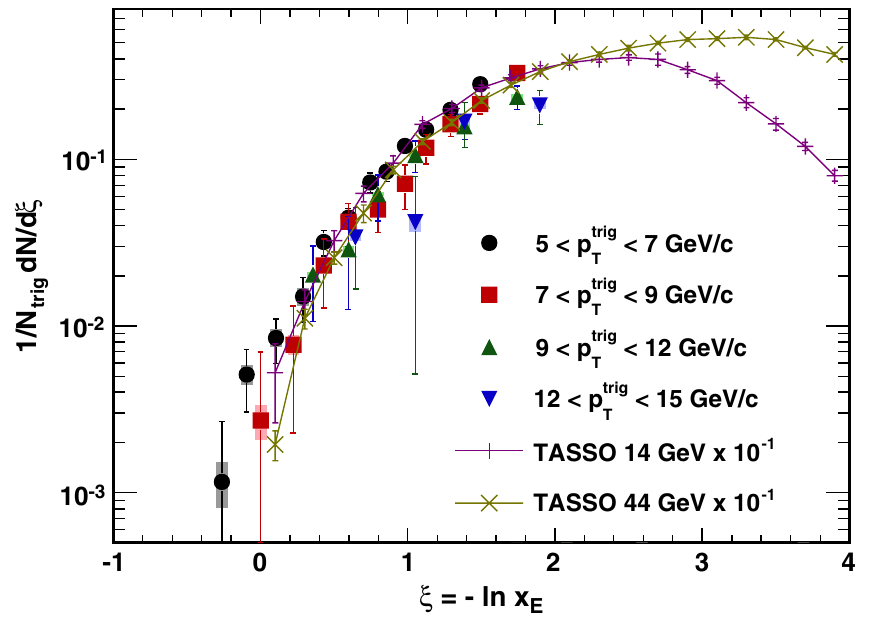} 
\caption[] {a) (left) PHENIX measurement~\cite{RTbook760}  of direct-$\gamma$-h correlations in p-p collisions at \sqs=200 GeV in the variable \mbox{$\xi=-\ln x_E\approx-\ln(p_T^{h}/p_T^{\gamma})$} compared to fragmentation functions measured in $e^+ e^-$ collisions at \sqs=14 and 44 GeV by TASSO~\cite{RTbook760}. b)(right)-(top) $\xi$ distributions of direct-$\gamma$-h correlations in Au+Au and p-p collisions at \sqs=200 GeV. c) (right)-(bottom) Ratio of the Au+Au/p-p distributions, $I_{AA}(\xi)$ when the away side azimuthal range is restricted as indicated~\cite{PXPRL111gamh}}\hspace*{2pc}
       \label{fig:PXgamma-h} \end{minipage}}
\raisebox{0.0pc}{\begin{minipage}[b]{0.49\linewidth} \includegraphics[width=\linewidth]{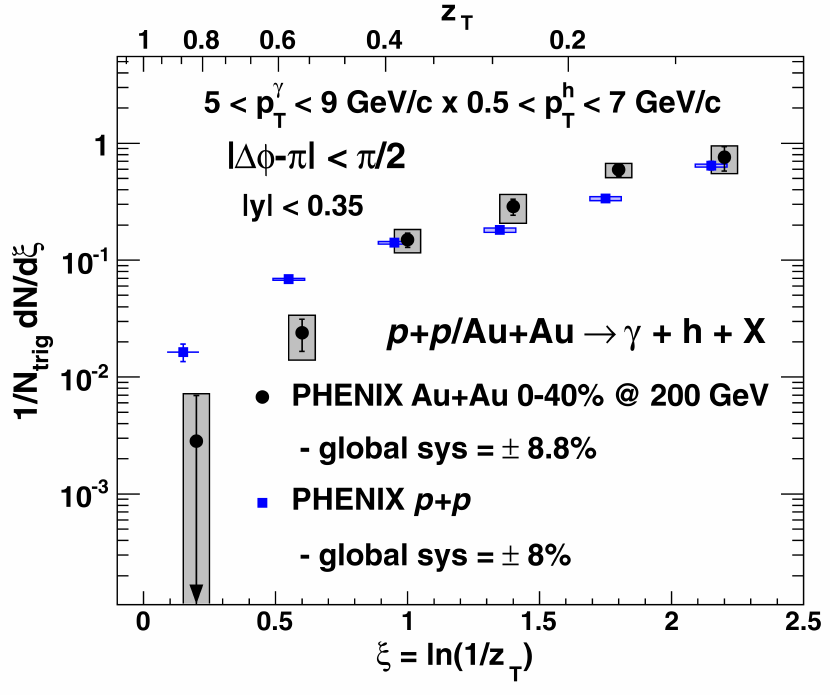} 
\includegraphics[width=\linewidth]{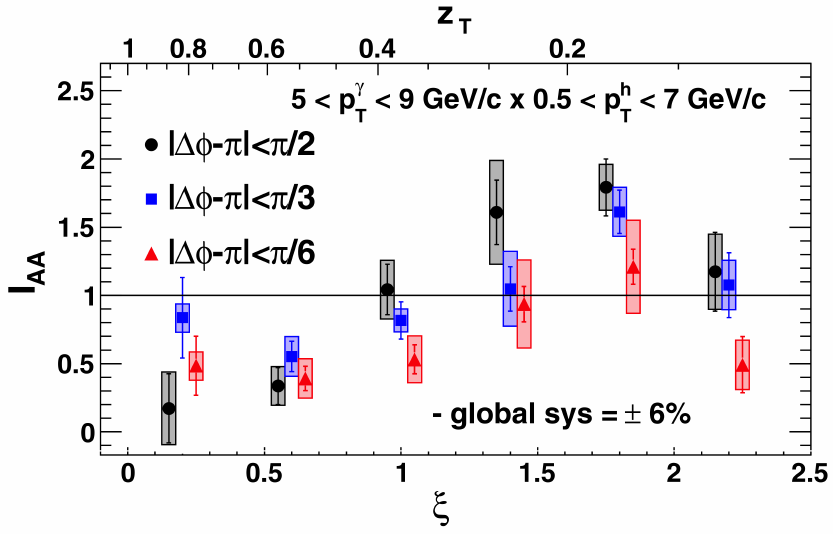} \end{minipage}}
   \end{figure}

In distinction to the case of di-hadron correlations (where both hadrons are fragments of jets) which do not measure the fragmentation function, direct-$\gamma$-hadron correlations do measure the fragmentation function of the jet from the away parton (mostlikely a $u$ quark~\cite{RTbook}). PHENIX has shown that this is true for p-p collisions Fig.~\ref{fig:PXgamma-h}a~\cite{RTbook760}. This year, improved measurements by PHENIX~\cite{PXPRL111gamh} in both p-p and Au+Au collisions (Fig.~\ref{fig:PXgamma-h}b) now indicate a significant modification of the fragmentation function in Au+Au (0-40\%) central collisions compared to p-p , with an enhancement at low $z_T=p^{h}_{T}/p^{\gamma}_{T}$ (large $\xi=-\ln z_T$)  and a suppression at large $z_T$ (small $\xi$)  which is more clearly seen as $I_{AA}(\xi)$, the ratio of the fragmentation functions in Au+Au/pp (Fig.~\ref{fig:PXgamma-h}c). As shown in Fig.~\ref{fig:PXgamma-h}c, restricting the away-side azimuthal range reduces the large $\xi>0.9$ ($p^{h}_{T}\lsim 3$ GeV/c) enhancement but leaves the suppression at small $\xi<0.9$  relatively unchanged, which shows that the large $\xi$ enhancement is predominantly at large angles, similar to the effect observed by CMS with actual jets.~\cite{CMSQM2012}. 

\begin{figure}[!h] 
      \centering
       \includegraphics[width=0.49\linewidth]{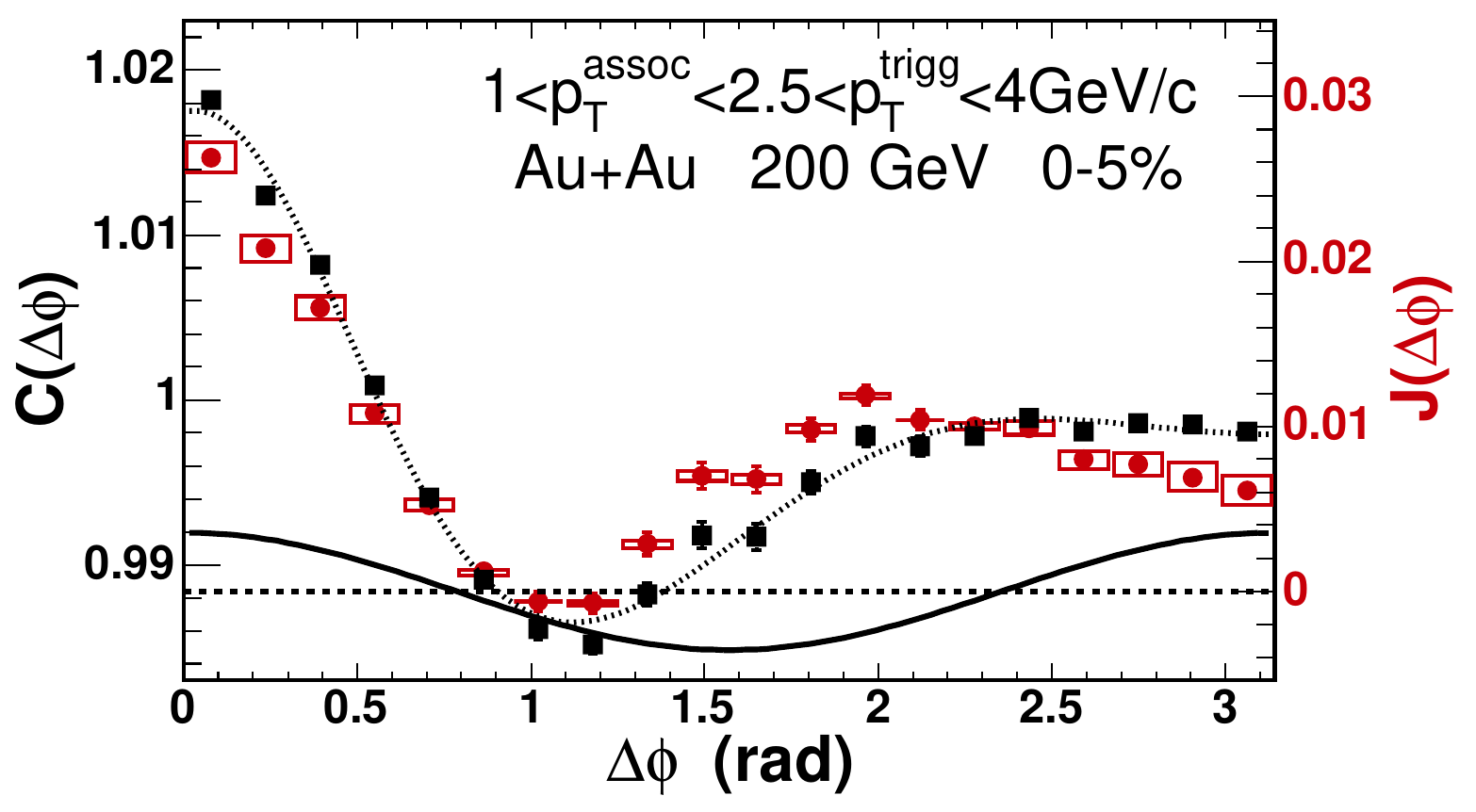}\hspace{1pc}
\raisebox{0.0pc}{\includegraphics[width=0.42\linewidth]{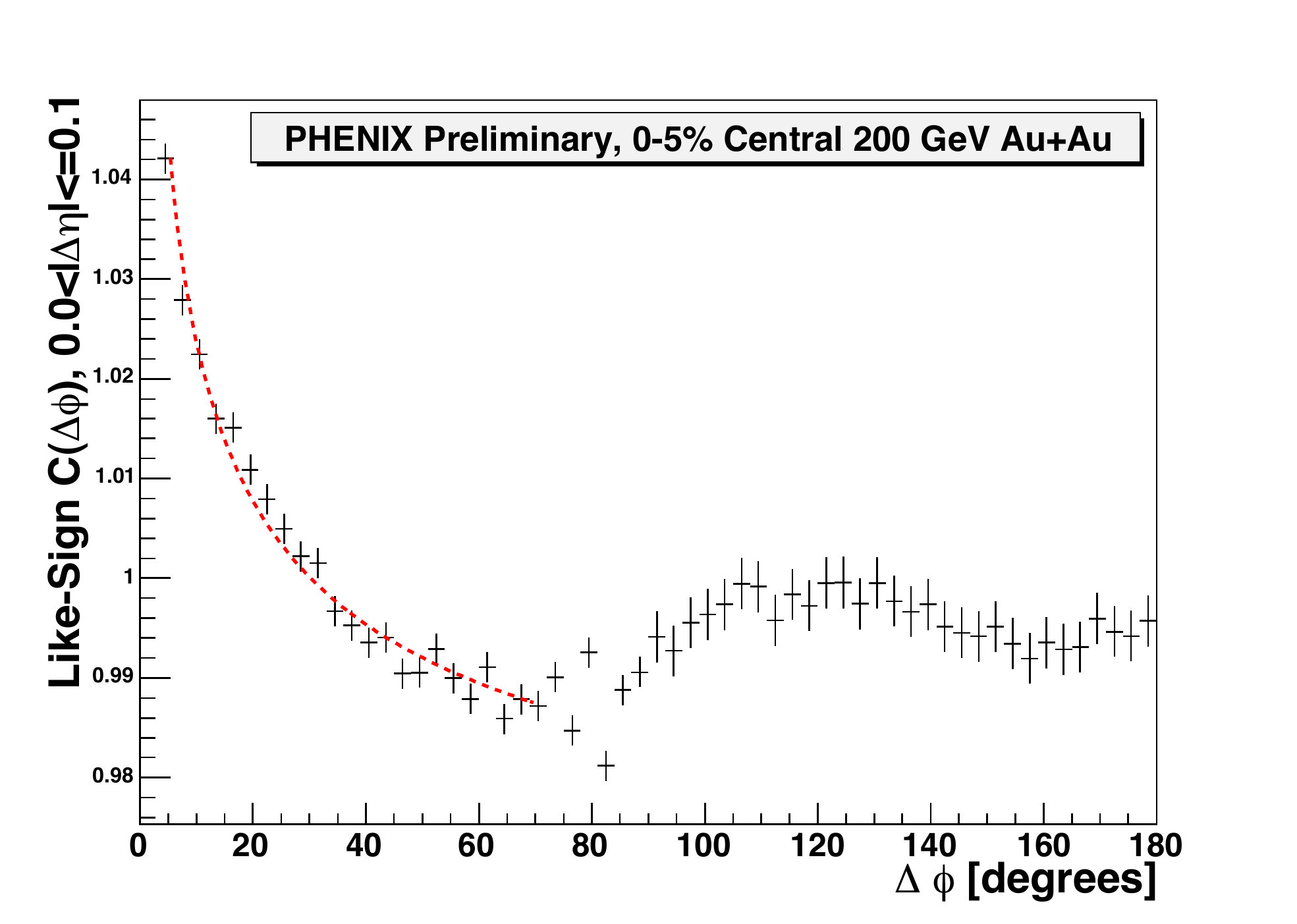}}%%{ij23b}       
%\raisebox{0.0pc}{\includegraphics[width=0.41\linewidth]{figs/JTM-cpod09-wMJTannotations}}%%{ij23b}
\caption[] {a) (left) Azimuthal correlation $C(\Delta\phi)$ (\fullsquare\;) of $h^{\pm}$ with $1\leq p_{T_a}\leq 2.5$ GeV/c from a ``high $p_T$'' trigger $h^{\pm}$ with $2.5\leq p_{T_t}\leq 4$ GeV/c in Au+Au central (0-5\%) collisions at $\sqrt{s_{NN}}=200$ GeV~\cite{PXhpairPRL98}. The solid line is the correction for $v_2$ which is subtracted to give the jet correlation function $J(\Delta\phi)$ ({\color{Red}\;\fullcircle}); b)(right) $C(\Delta\phi)$ for like sign $h^{\pm}$-pairs with $0.2<p_{T_1}, p_{T_2} <0.4$ GeV/c for both particles~\cite{JTMQM06}.} 
      \label{fig:byebye}
   \end{figure}

The principal difficulty in observing the broadening of di-jet or di-hadron azimuthal correlations by the transport coefficient $\hat{q}$ of the \QGP\ stems from artifacts with names such as ``Mach Cone'', ``Ridge'', ``Head and Shoulders'' (Fig.~\ref{fig:byebye}). Perhaps one should have been suspicious because the artifact was observed in correlations triggered by both high (Fig~\ref{fig:byebye}a) and low (Fig~\ref{fig:byebye}b) $p_T$  particles. 
      %\vspace*{-0.05in}
These artifacts are now known to be due to odd harmonics of collective flow, notably $v_3$, which were thought to be forbidden by the assumed azimuthal symmetry $\phi\rightarrow \phi +\pi$ of the almond shaped participant region of an A+A collision but were shown (only recently~\cite{AlverRoland}) to be produced from event-by-event fluctuations in the collision geometry.

\begin{figure}[!h] 
      \centering
\raisebox{0.0pc}{\includegraphics[width=0.45\linewidth]{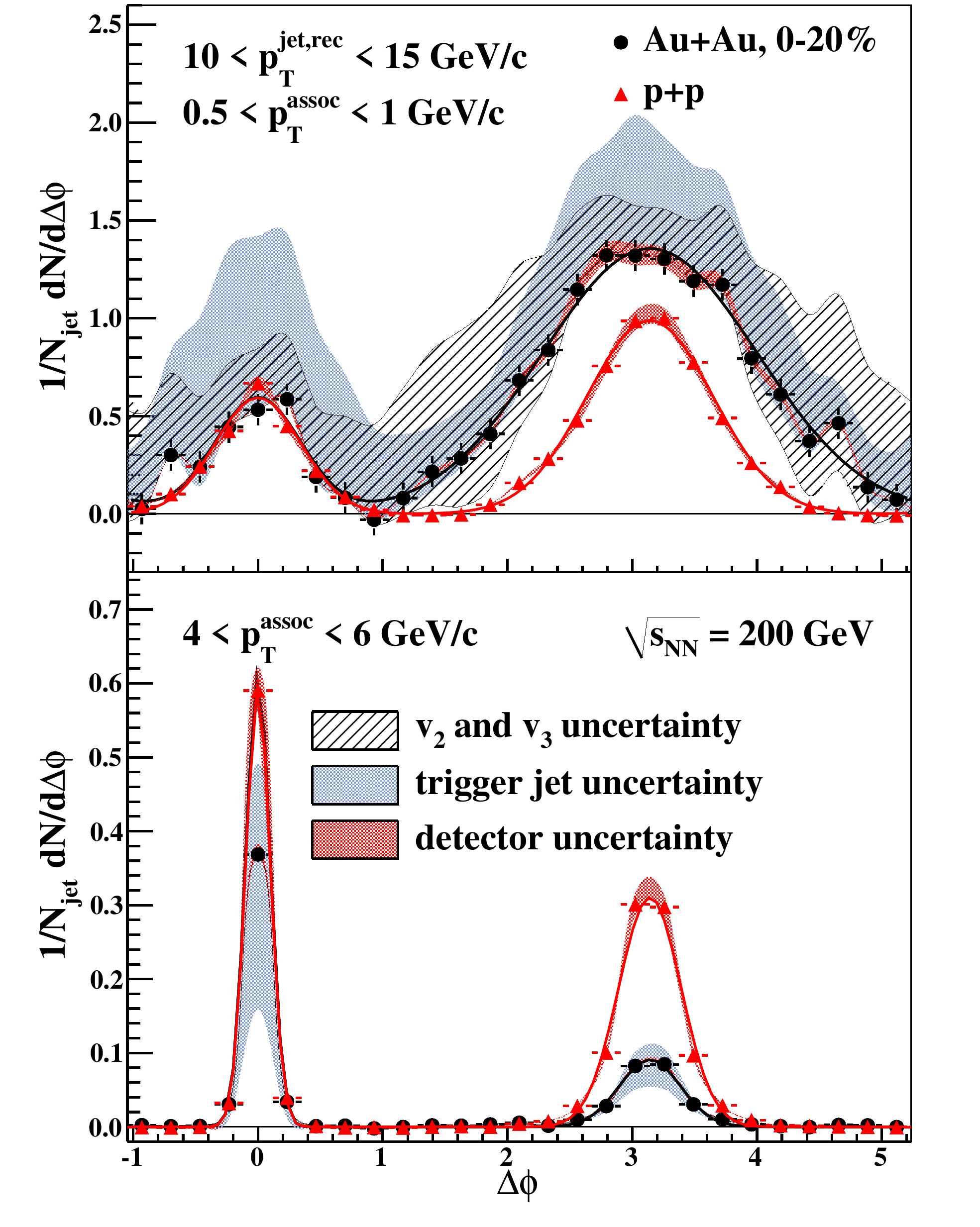}} 
\raisebox{0.0pc}{\begin{minipage}[b]{0.42\linewidth}\includegraphics[width=\linewidth]{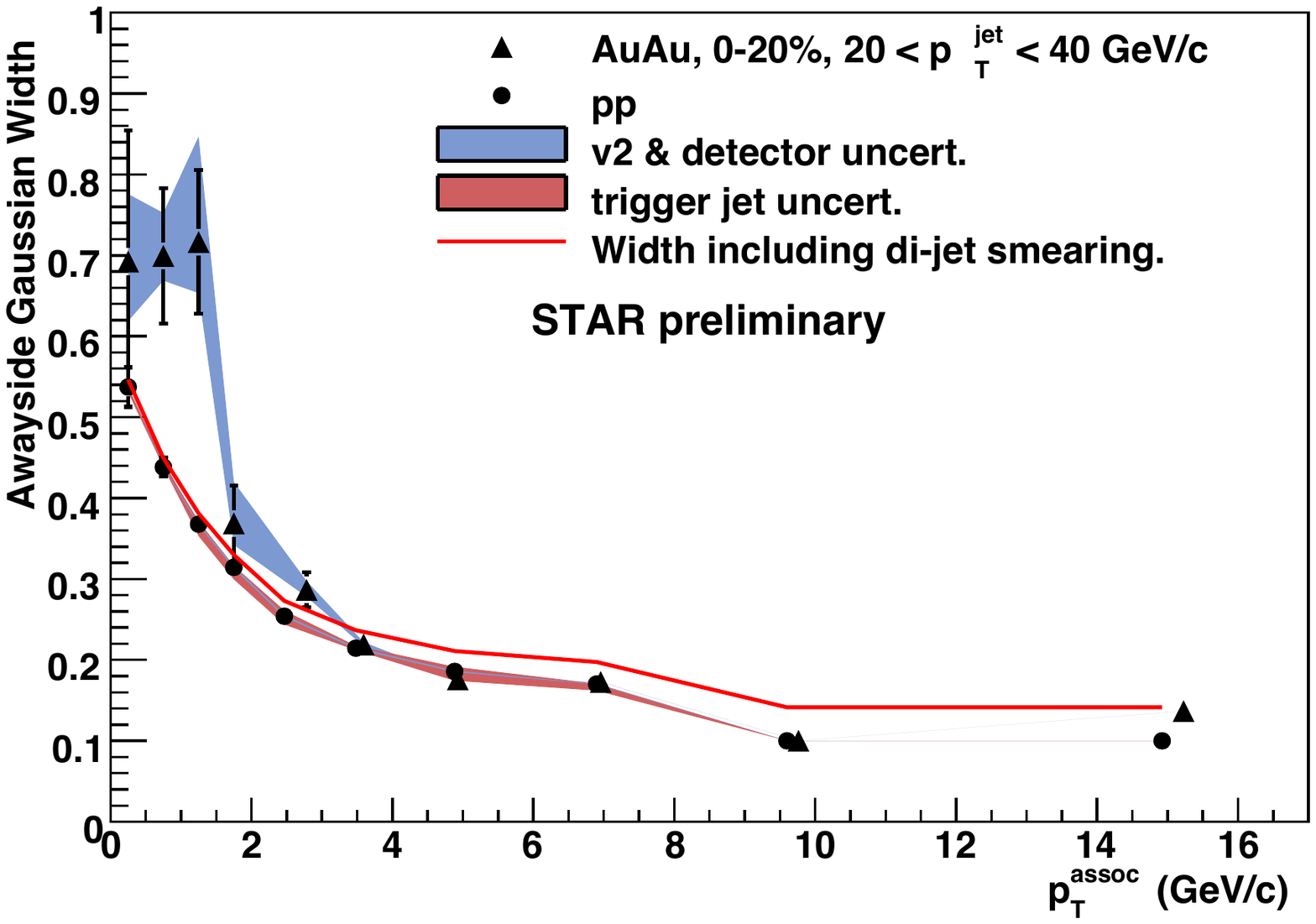}
\includegraphics[width=\linewidth]{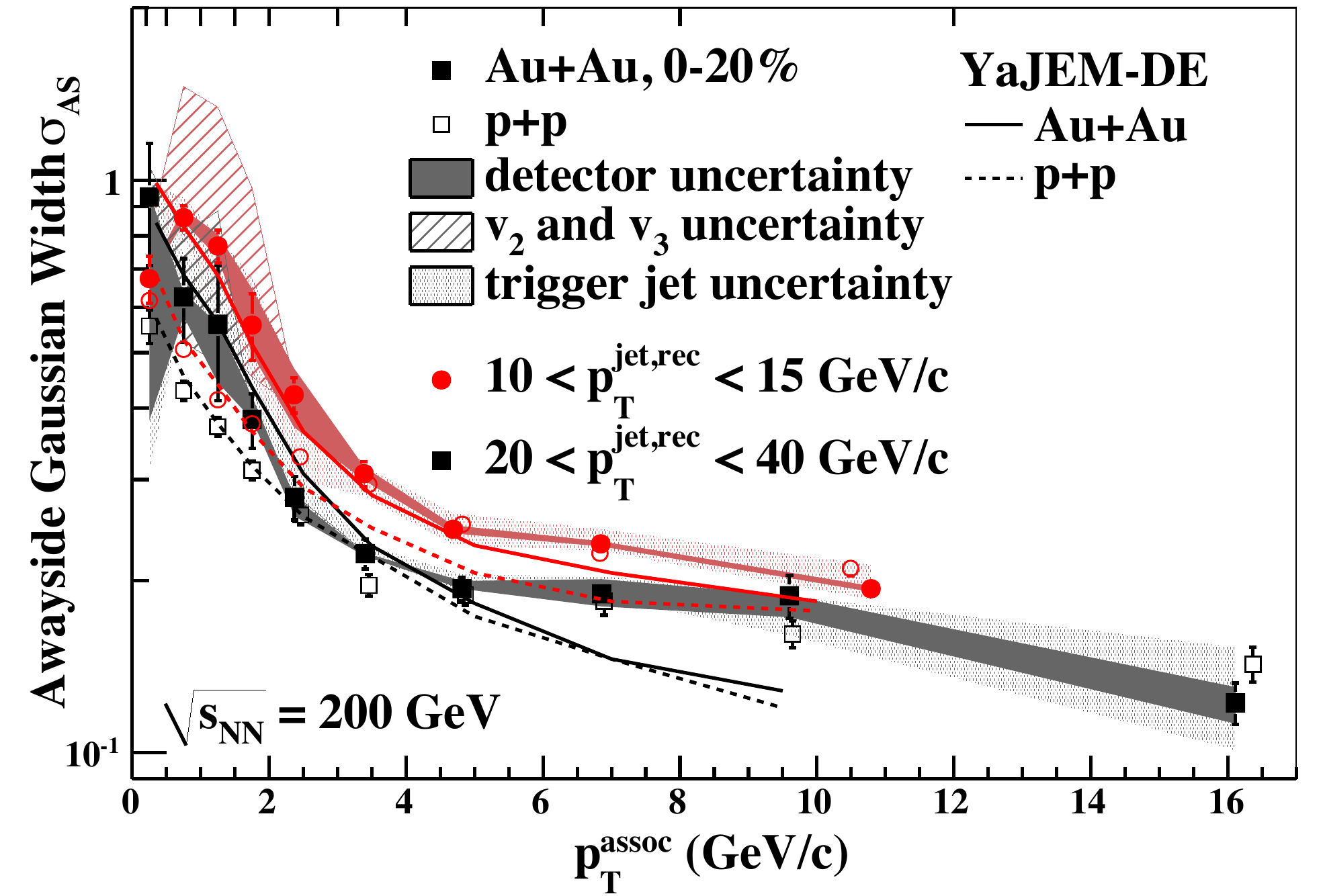} \end{minipage}}
\caption[] {a) (left) Azimuthal correlation $J(\Delta\phi)$ with systematic uncertainties shown~\cite{STARJet-h-pub}. b)(right)-(top) Preliminary result~\cite{CainesQM11} for Awayside rms width, $\sigma_{\rm AS}$, as a function of $p_{T_a}$. c)(right)-(bottom) Final result~\cite{STARJet-h-pub} for $\sigma_{\rm AS}$, as a function of $p_{T_a}$.} 
      \label{fig:Caines}
   \end{figure}

Of course, understanding that the extra ``bumps'' in the correlation function are due to odd harmonics still requires one to know the values of these harmonics in order to subtract them. This is still the largest systematic uncertainty in attempts to observe the $\hat{q}$-broadening, for instance in a recent measurement by STAR of jet-hadron correlations~\cite{STARJet-h-pub} (Fig.~\ref{fig:Caines}). 
       %\vspace*{-0.05in}
When the full systematic uncertainties, including those on $v_2$ and $v_3$ (Fig.~\ref{fig:Caines}a), are taken into account,  the result for the medium induced broadening of the away-side widths, $\sigma_{\rm AS}$, in Au+Au relative to p-p which looked significant in the preliminary results (Fig.~\ref{fig:Caines}b)~\cite{CainesQM11} become only ``suggestive of medium-induced broadening~\cite{STARJet-h-pub}'' in the final result (Fig.~\ref{fig:Caines}c) because ``they are highly dependent on the shape of the subtracted background~\cite{STARJet-h-pub}'', notably the $v_2$ and $v_3$ of the trigger jets. 

\begin{figure}[!h] 
      \centering
       \includegraphics[width=0.5\linewidth]{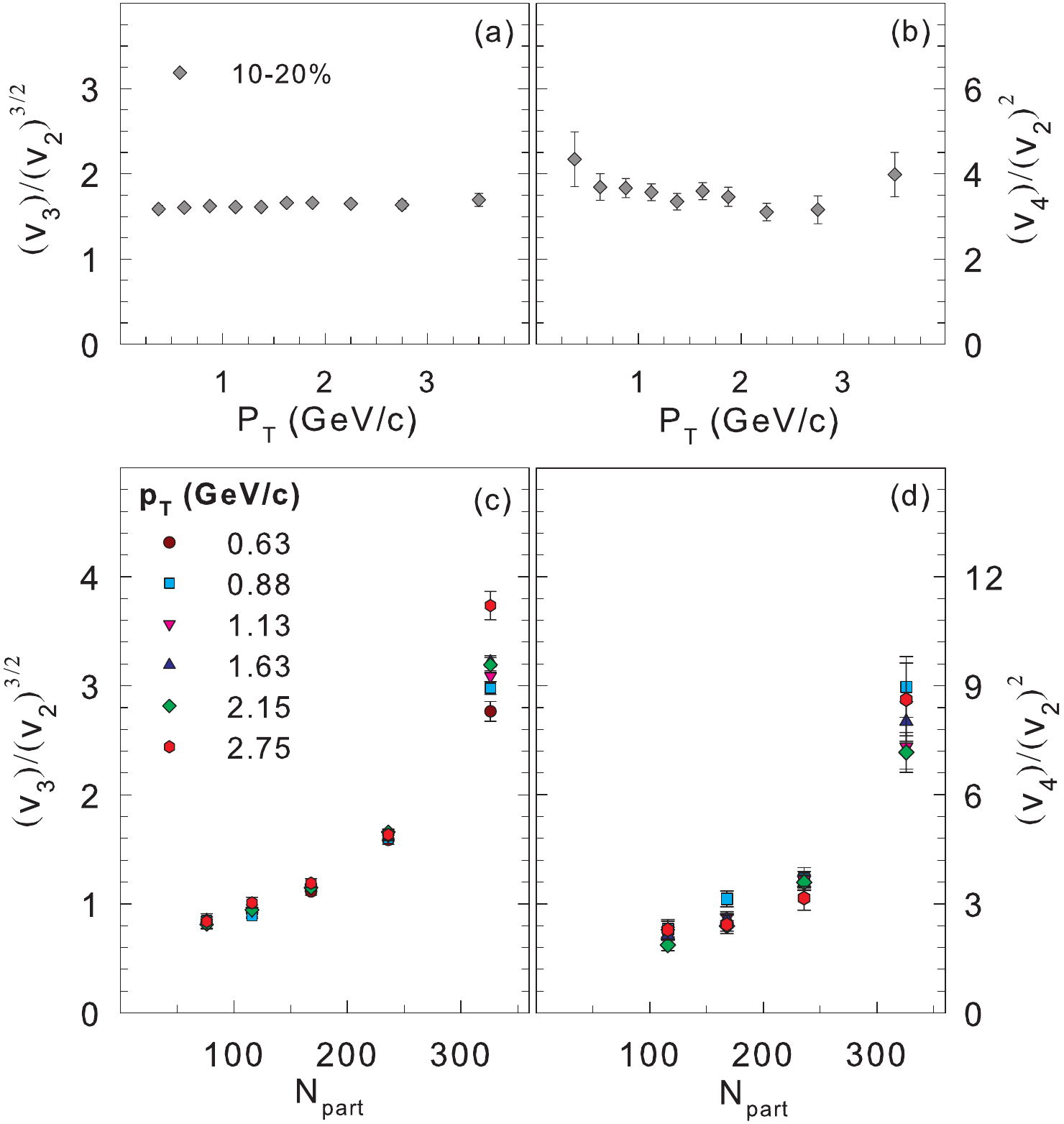}\hspace{2pc}
\raisebox{3pc}{\begin{minipage}[b] {0.4\linewidth}        
\caption[] {a) $v_3/(v_2)^{3/2}$ vs. $p_T$ for 10-20\% central Au+Au collisions at \sqsn=200 GeV; b) same for $v_4/v_2^2$; c) $v_3/(v_2)^{3/2}$ vs centrality (\Npart) for several $p_T$ cuts; d) same for $v_4/v_2^2$. Plot is from Ref.~ \cite{LaceyScaling2011} using data from Ref.~\cite{PXvnPRL107}.}
      \label{fig:acoustic}\end{minipage}}
   \end{figure}

One way out of this dilemma, at least for di-hadron correlations, has been pointed out by Roy Lacey and his group~\cite{LaceyScaling2011} who proposed ``acoustic scaling'' of the harmonics in analogy with attenuation of pressure driven sound waves in a viscous medium. They observe that viscous corrections damp the eccentricity-driven flow harmonics, $v_n$ with increasing $n$ but have minimal effect on the $p_T$ dependence because of the small values of $\eta/s$. The proposed ``acoustic scaling'' that $v_n/(v_2)^{n-2}$ is independent of $p_T$ was verified, at least for $v_3$ and $v_4$ (Fig.~\ref{fig:acoustic})~\cite{LaceyScaling2011}, using previously published PHENIX measurements~\cite{PXvnPRL107}. 
      %\vspace*{-0.05in}
This opens the way to reduce the systematic uncertainty in subtracting the harmonic background to di-hadron correlations: once $v_2(p_T)$ is known or determined at any value of $p_T$ and centrality, the acoustic scaling can be used to constrain the higher harmonics. 

Another major topic this year was the p+Pb run at LHC which spurred new or improved d+Au results from RHIC. There is both clarity and confusion in the d+Au results. First, the clarity. Switching momentarily to soft multi-particle physics, new PHENIX measurements of \Et distributions at mid-rapidity in p-p, d+Au and Au+Au collisions at \sqsn=200 GeV verify that starting from the p-p distribution, the Au+Au distribution can be calculated by taking the fundamental elements of particle production as the number of constituent-quark participants, as proposed in 2003 ~\cite{EreminVoloshin} (NQP model). For symmetric systems such as Au+Au the NQP model is equivalent to the Additive Quark Model (AQM) from 1982~\cite{AQMPRD25} in which the fundamental element is a color string streched between constituent quark participants in the projectile and target with the additional restriction that only one color-string can be attached to a quark-participant. For asymmetric systems such as d+Au, the models differ because the number of color-strings is proportional only to the number of quark-participants in the smaller nucleus. 
         \begin{figure}[!h]
   \centering
\includegraphics[width=0.49\textwidth]{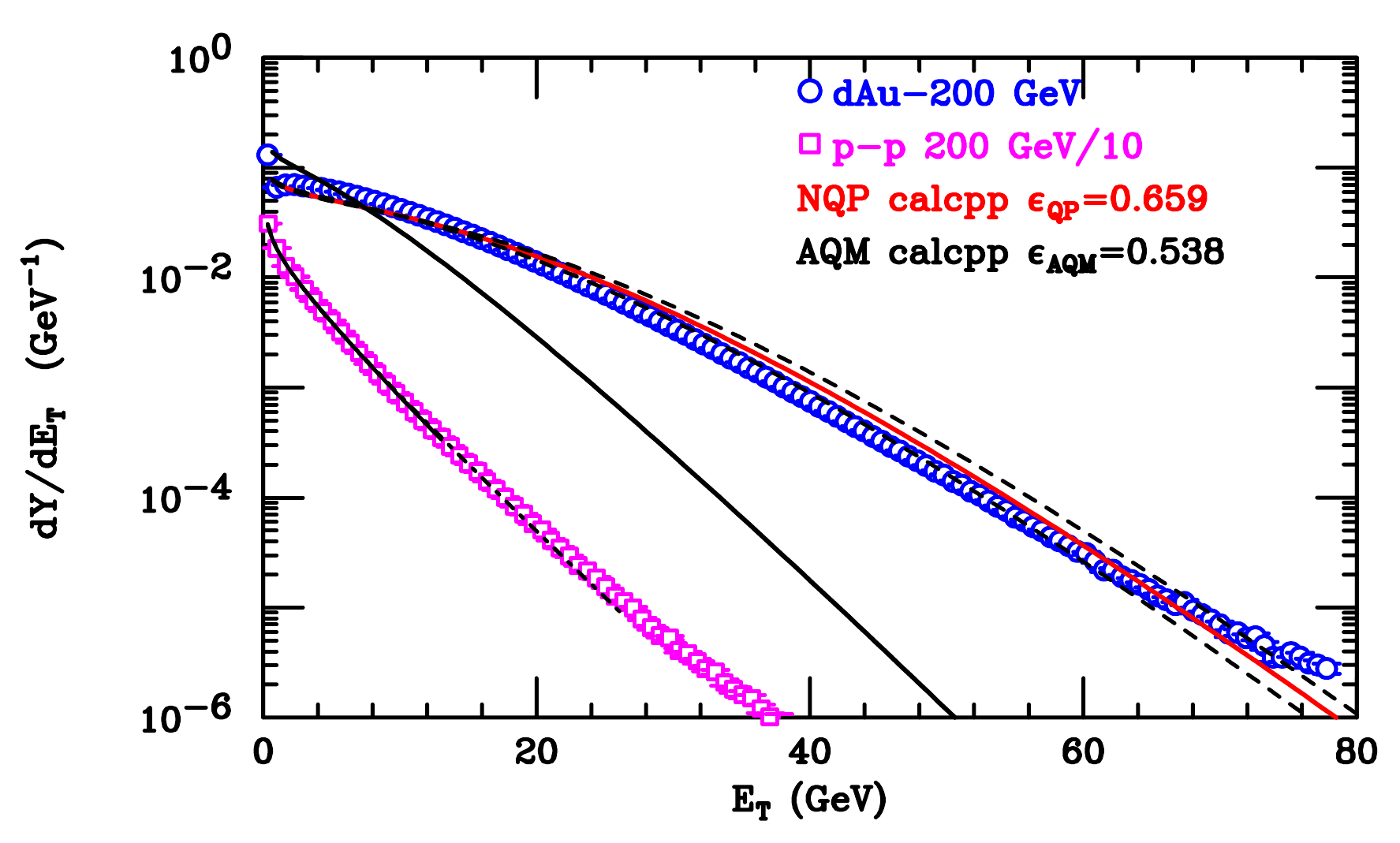}
\includegraphics[width=0.49\textwidth]{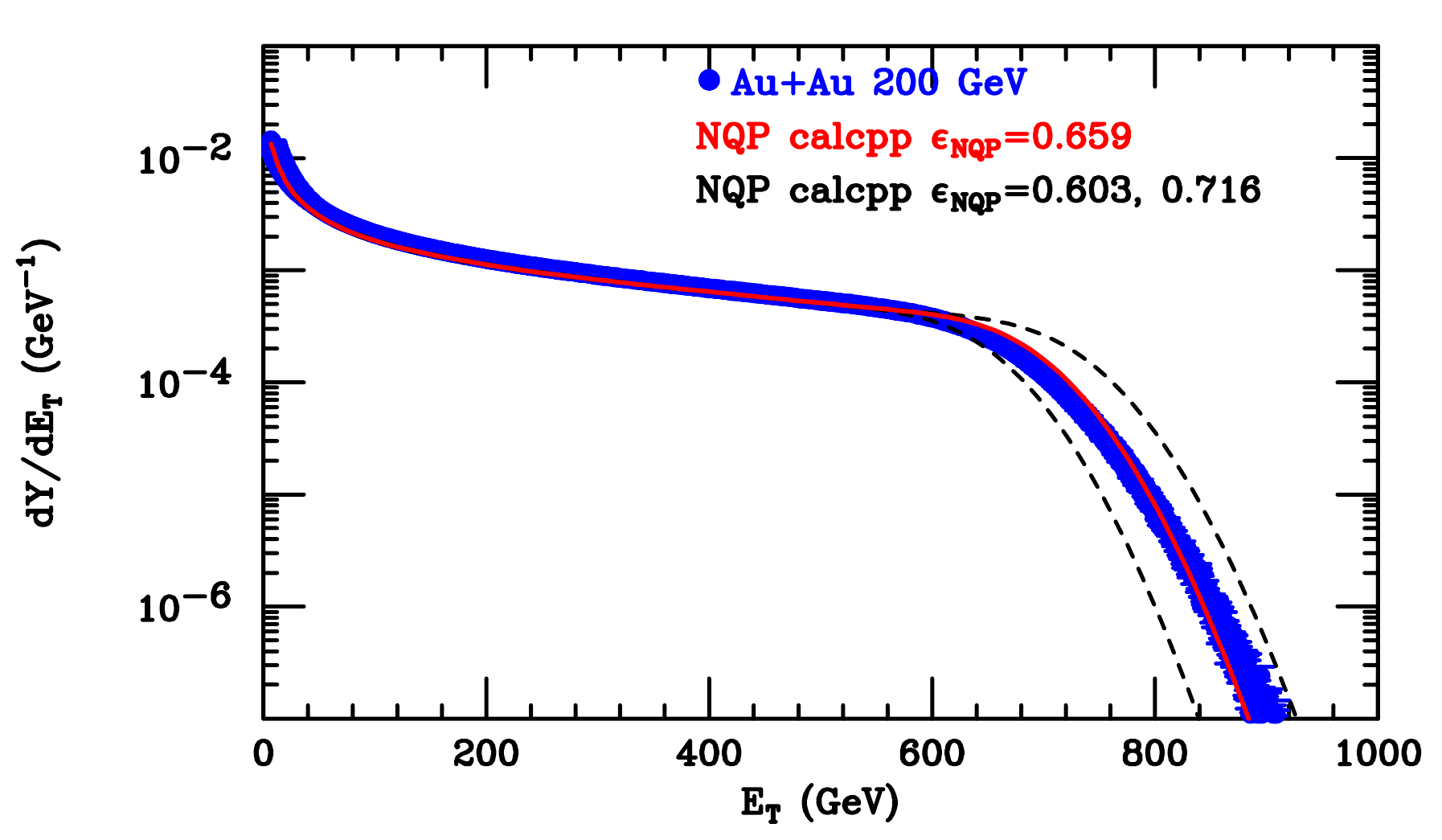}%\vspace*{-1.7pc}
\caption[]{a) (left) PHENIX measurement~\cite{ppg100} of mid-rapidity \Et distributions for p-p (\opensquare ) and d+Au ($\opencircle\!$) at \sqsn=200 GeV with calculations of the d+Au spectrum from fits to the p-p data (lines) based on the AQM (color-strings) and the number of constituent-quark participants (NQP). b) (right) Au+Au \Et distribution and NQP calculation. Systematic uncertainties are shown by dashed lines in both (a) and (b).  
\label{fig:PXNQP}}%\vspace*{-1.8pc}
\end{figure}
In Fig.~\ref{fig:PXNQP}~\cite{ppg100} the calculations which are based only on the measured p-p \Et distribution and Monte Carlo Glauber calculations of the nuclear geometry for the various cases show that the NQP calculation agrees with both the d+Au (Fig.~\ref{fig:PXNQP}a) and Au+Au (Fig.~\ref{fig:PXNQP}b) \Et measurements, but that the AQM exhibits a factor of 1.7 less \Et emission than observed in d+Au (Fig.~\ref{fig:PXNQP}a) due to the restriction on the number of effective constituent quark participants in the larger nucleus. In the Glauber calculation, the positions of the three constituent-quarks are generated about the position of each nucleon according to the measured charge distribution of the proton, which gives a physical basis for ``proton size fluctuations'' in nuclear geometry calculations. 

\begin{figure}[!h]
   \centering
\includegraphics[width=0.62\textwidth]{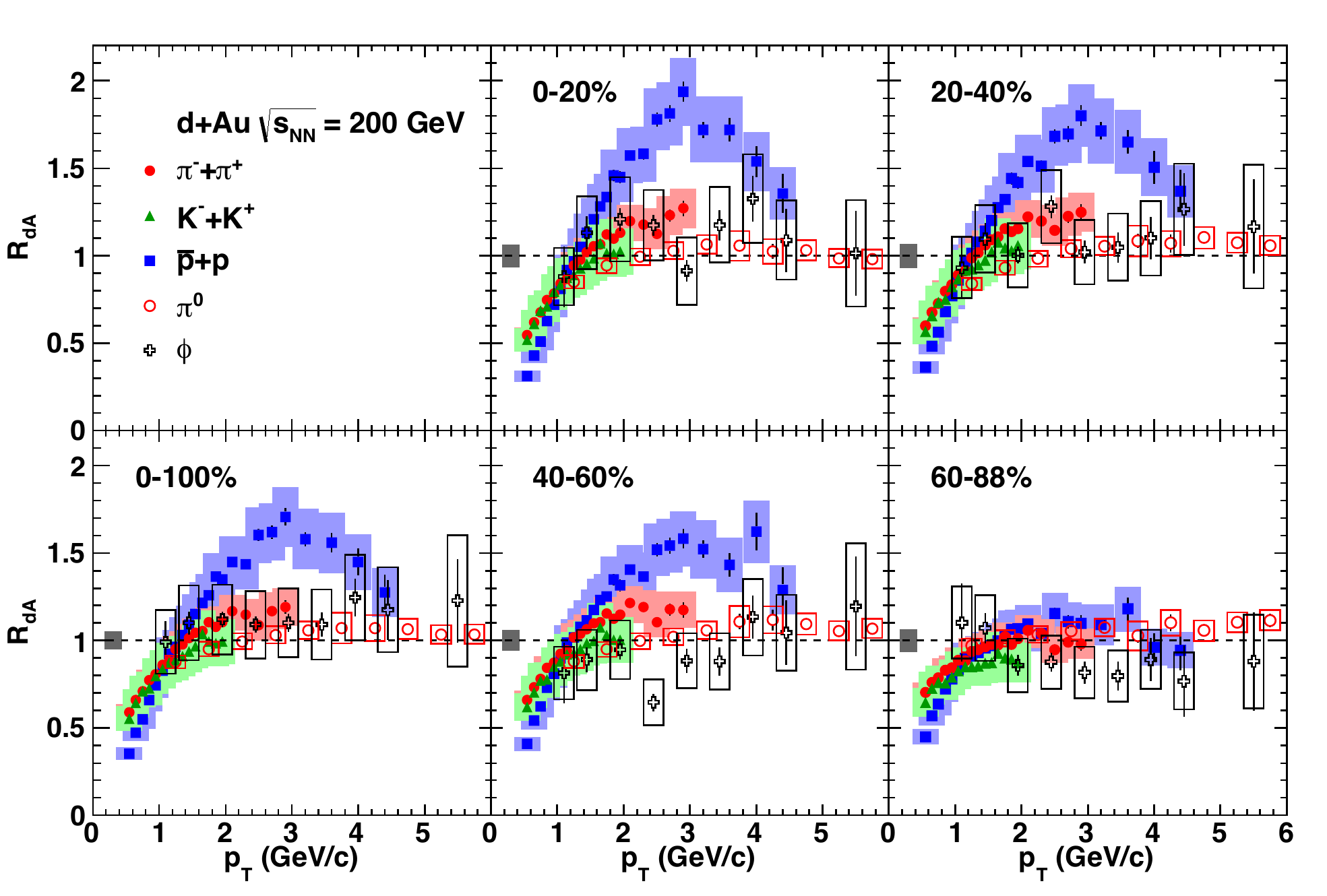}\hspace{0.3pc}
\includegraphics[width=0.36\textwidth]{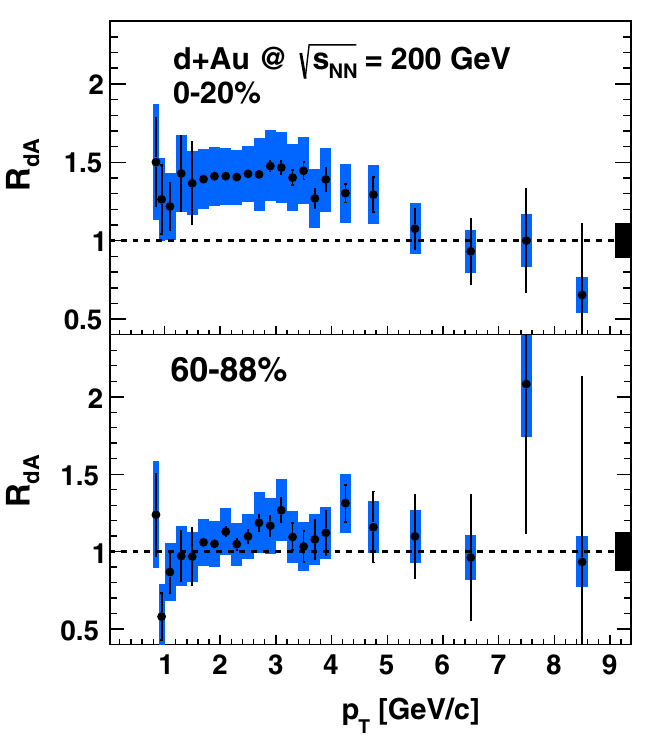}%\vspace*{-1.7pc}
\caption[]{a) (left) PHENIX measurement~\cite{ppg146} of $R_{\rm dA}$ of identified particles as a function of \pT and centrality in d+Au collisions at \sqsn=200 GeV.  b) (right) PHENIX measurement~\cite{PXeHF-dAuPRL109} of $R_{\rm dA}$ for direct-single $e^{\pm}$ from Heavy Flavor ($c$ and $b$ quark) decay for peripheral (60-88\%) and central (0-20\%) d+Au collisions at \sqsn=200 GeV. 
\label{fig:newdAuPX}}%\vspace*{-1.8pc}
\end{figure}

On the hard-scattering front, new results for identified charged hadrons in d+Au (Fig.~\ref{fig:newdAuPX}a)~\cite{ppg146}  may give a clue for understanding the baryon anomaly. 
In all centrality classes, the mesons, $\pi^0$, $\pi^{\pm}$, $K^{\pm}$ and $\phi$ show point-like behavior, $R_{\rm dA}=1$, over the range $2\leq p_T\leq 6$ GeV/c, while, apart from the most peripheral bin, the protons exhibit a huge enhancement (Cronin effect~\cite{CroninEffect}) for $2\leq p_T\leq 4$ GeV/c similar to the baryon anomaly in Au+Au (recall Fig.~\ref{fig:allPXpid}). In fact, as originally shown in a previous PHENIX measurement~\cite{ppg30} and confirmed with the new more precise measurements~\cite{ppg146}, the proton enhancement in d+Au is generally considerably larger than in Au+Au. The  new observation makes it clear that a common explanation of the d+Au and Au+Au baryon enhancements for $2\leq p_T\lsim 4-6$ GeV/c is needed. 

    It has been popular, though not necessarily correct~\cite{PXPRC71R,PXPLB649}, to attribute the baryon anomaly to recombination of soft partons rather than fragmentation of hard-scattered partons~\cite{FriesMuellerPRL90,HwaYangdAuPRC70} because the suppression has been observed to be related to a ``baryon vs. meson dynamic, as opposed to a simple mass-dependence~\cite{ppg146}''. However a new d+Au measurement of single $e^\pm_{\rm HF}$ from $c$ and $b$ heavy quarks, mostlikely from the decay of $D$-mesons with mass twice that of the proton, appears to create another layer of doubt to  the recombination argument (Fig.~\ref{fig:newdAuPX}b~)\cite{PXeHF-dAuPRL109}. In Au+Au (recall Fig.~\ref{fig:allPXpid}b), the $e^\pm_{\rm HF}$ are suppressed like the ($K^+$) mesons in the range $2\leq p_T\leq 6$ GeV/c; but in d+Au (Fig.~\ref{fig:newdAuPX}b) they show a Cronin enhancement like the baryons, not like the mesons, in central (0-20\%) collisions. Thus the recombination issue is coupled to understanding the Cronin effect~\cite{CroninEffect} which suffers from the problem that in the nearly 40 years since its discovery, the exact cause of the Cronin effect still remains a mystery. 

\begin{figure}[!h] 
      \centering
\raisebox{0.0pc}{\begin{minipage}[b]{0.42\linewidth}\includegraphics[width=\linewidth]{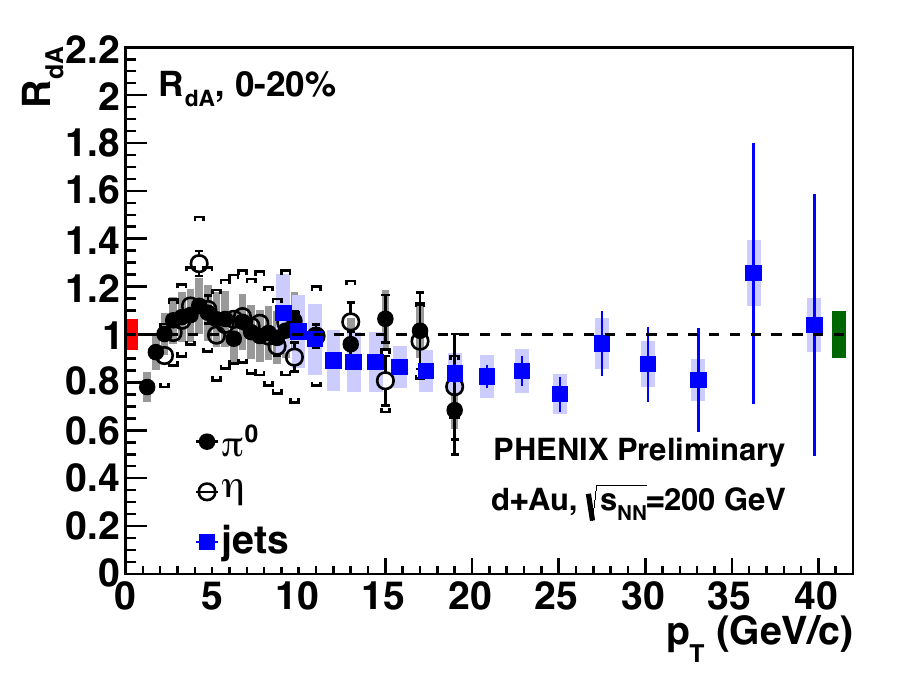}
\includegraphics[width=\linewidth]{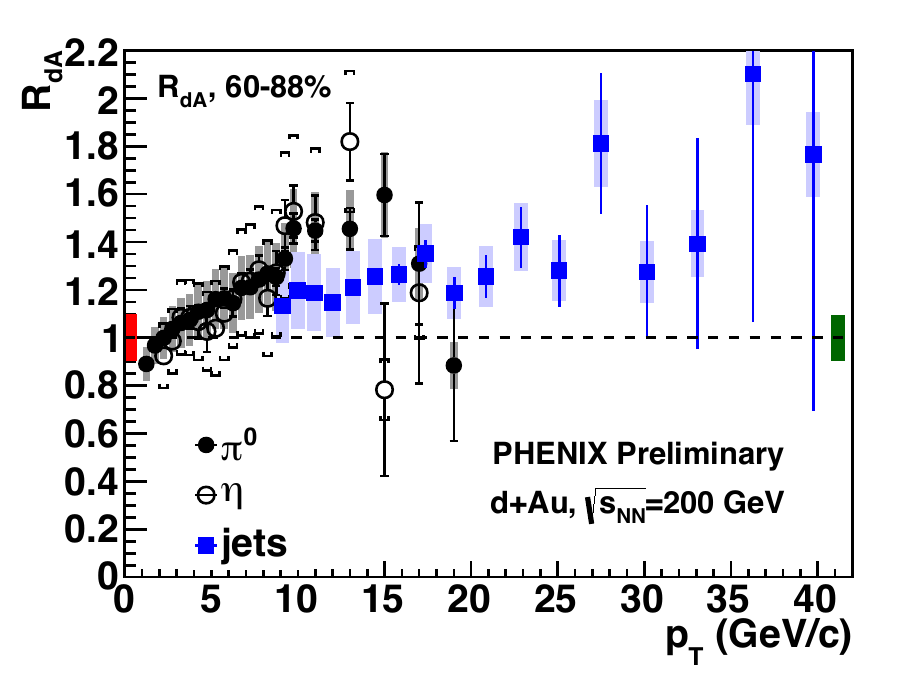} \end{minipage}}
\raisebox{0.0pc}{\includegraphics[width=0.46\linewidth]{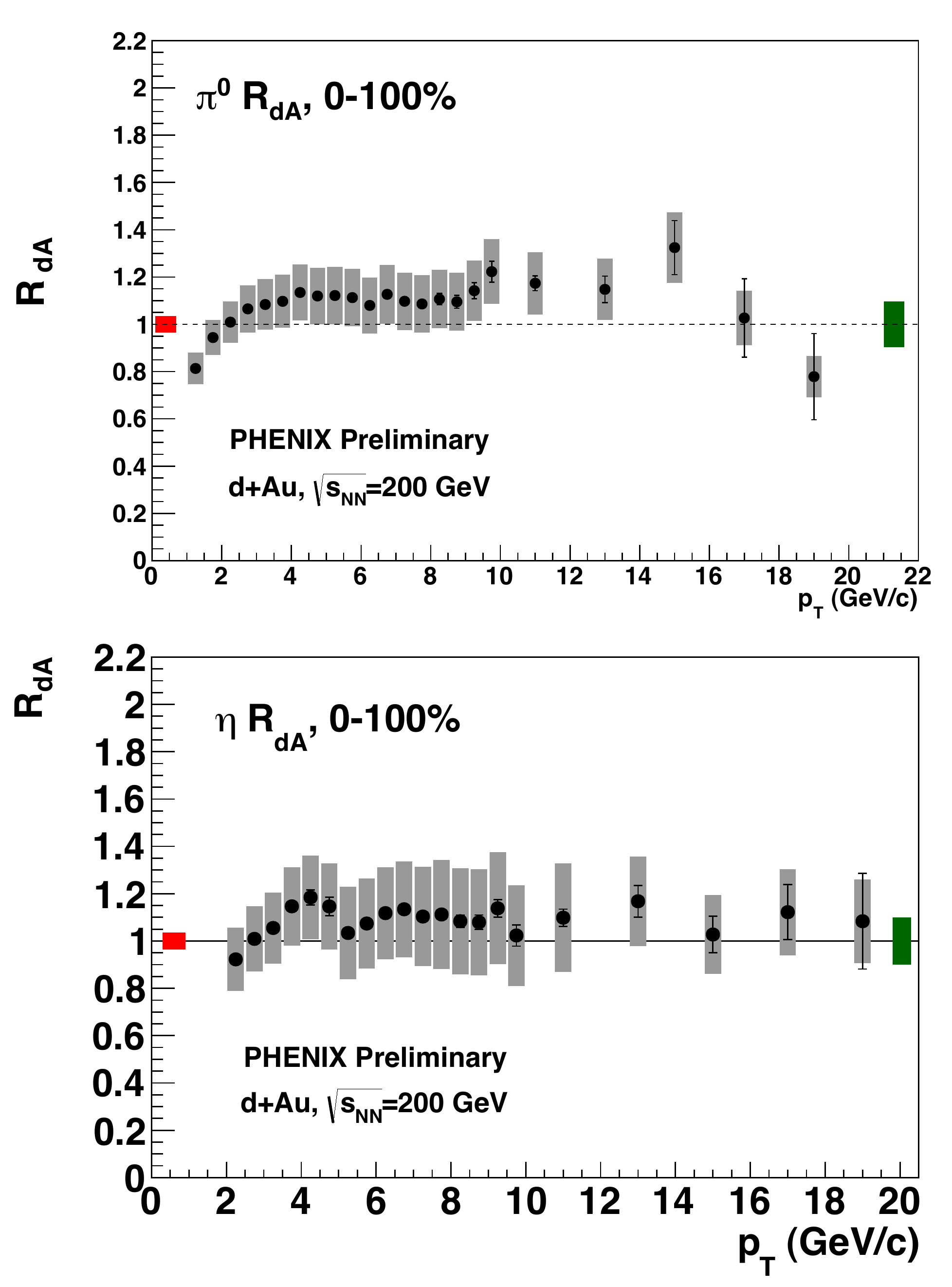}} 
\caption[] {a) (left) $R_{\rm dA}$ for $\pi^0$, $\eta$ and jets as a function of \pT for central (0-20\%) and peripheral (60-88\%) d+Au collisions at \sqsn=200 GeV~\cite{BaldoQM2012}. b)(right) $R_{\rm dA}$ for minimum bias $\pi^0$ and $\eta$~\cite{BaldoQM2012}. } 
      \label{fig:jetpi0etadAu}
   \end{figure}

Another confusing cold nuclear matter (CNM) effect concerns the definition of centrality in d+Au collisions. This is visible mostly in peripheral collisions for $p_T>10$ GeV/c. In Fig.~\ref{fig:newdAuPX}, the $R_{\rm dA}$ for peripheral (60-88\%) collisions for $2\leq p_T \leq 6$ GeV/c looked reasonable, i.e. $R_{\rm dA}\approx 1$ for all the particles, while in Fig.~\ref{fig:jetpi0etadAu}a~\cite{BaldoQM2012}, for higher $p_T$ $\pi^0$, $\eta$ mesons as well as jets , there appears to be a problem or something exciting according to one's taste. 
       %\vspace*{-0.05in}
The peripheral collisions show a huge enhancement (which doesn't make sense to the author because true peripheral collisions are larglely single nucleon-nucleon collisions) while the central collisions don't look unreasonable. Personally, I think that we should only use minimum bias p+A collisions to understand the nuclear-size dependence of any CNM effect because the hard-scattering cross section in p+A collisions is simply $A$ times the p-p cross section (modulo some possible isotopic spin effects, which are small). Glauber calculations and matching to some measure of centrality are not necessary.  Also the comparison with parton distribution functions measured in e+A collisions is straightforward because centrality has generally not been measured in electron scattering and is likely not measurable. In fact, the minimum bias $R_{\rm dA}$ for $\pi^0$ and $\eta$ from the same measurement~\cite{BaldoQM2012} (Fig.~\ref{fig:jetpi0etadAu}b) look reasonable; and the minimum bias jet $R_{\rm dA}$ looks similar to the minimum bias $\pi^0$  
$R_{\rm dA}$ rather than to the central collision jet $R_{\rm dA}$~\cite{DVPthesis}. A p+A run with several $A$ is planned at RHIC for 2015-16 in addition to runs with $^3$He+Au and possibly more d+Au. 

\begin{figure}[!t]
   \centering
\includegraphics[width=0.48\textwidth]{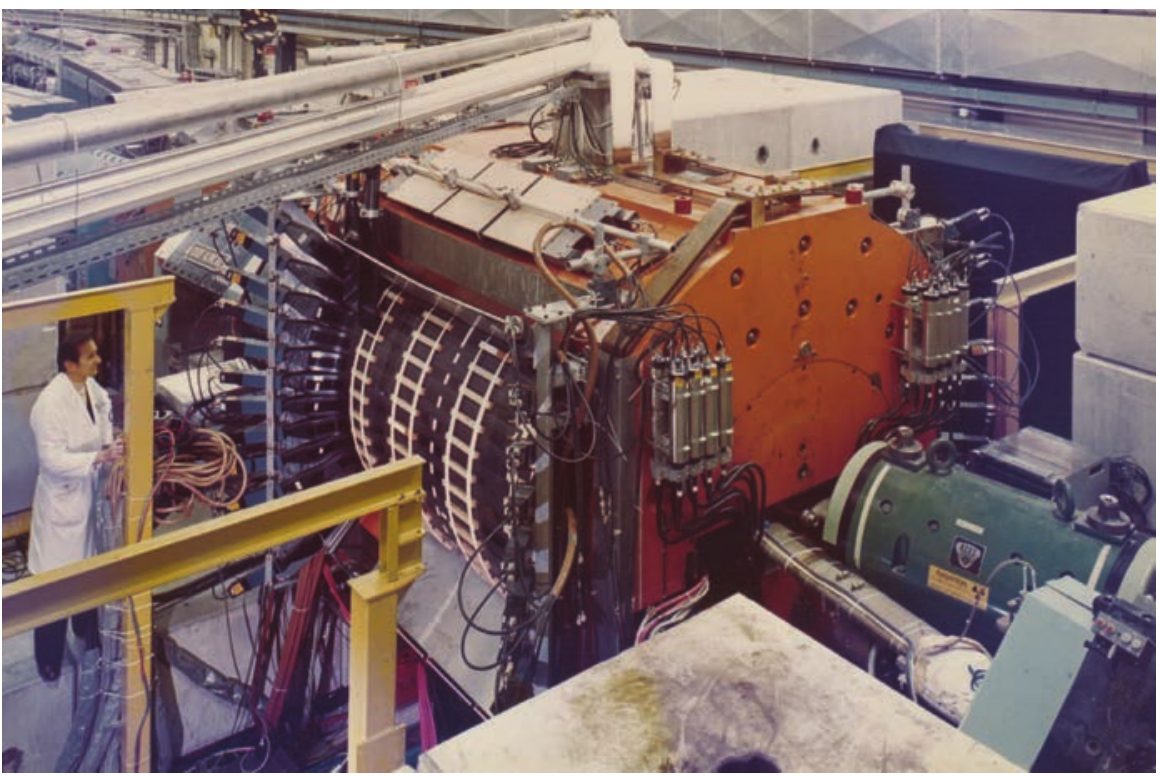}\hspace{0.3pc}
\includegraphics[width=0.48\textwidth]{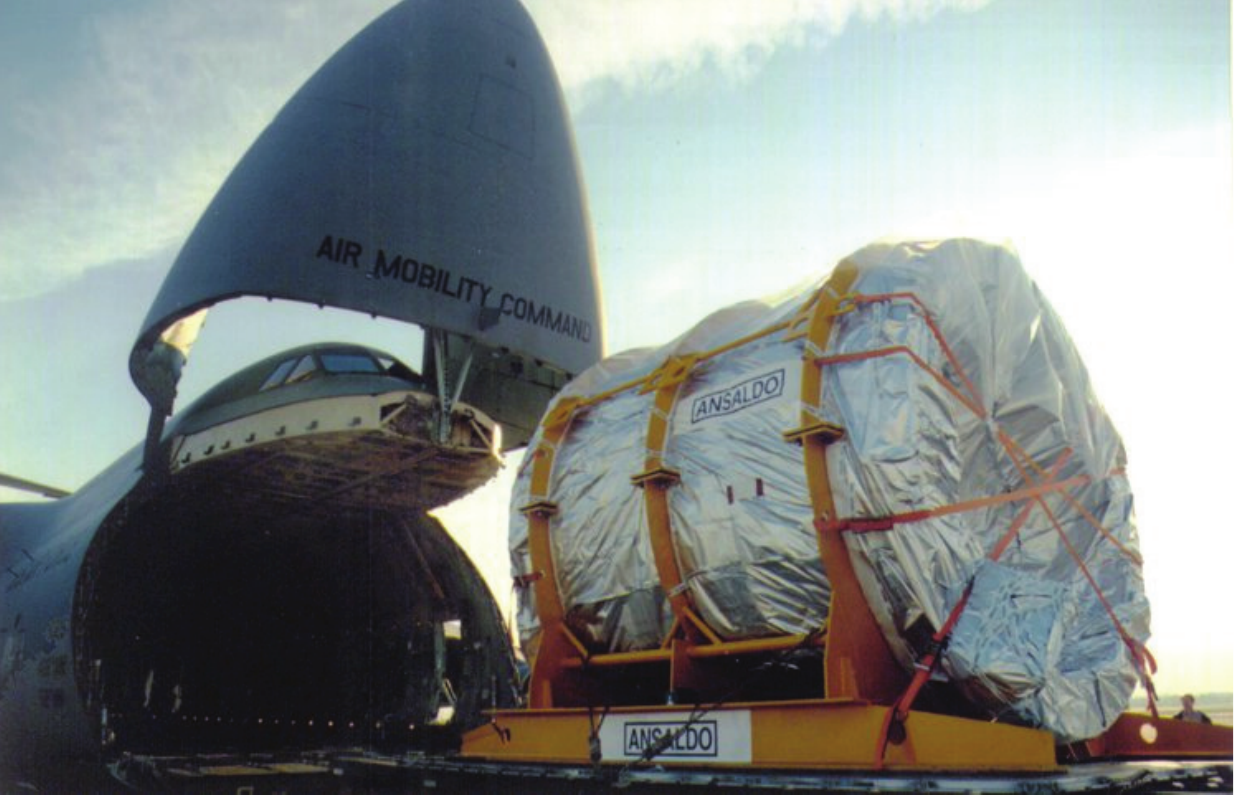}%\vspace*{-1.7pc}
\caption[]{a) (left) CCOR thin coil superconducting solenoid, 70 cm radius, installed at the CERN ISR in 1977. b) (right) BABAR thin coil superconducting solenoid, 1.5 m radius, being shipped from Ansaldo in Italy to SLAC in 1997.  
\label{fig:solenoids}}%\vspace*{-1.8pc}
\end{figure}

\begin{figure}[!h]
   \centering
\includegraphics[width=0.95\textwidth]{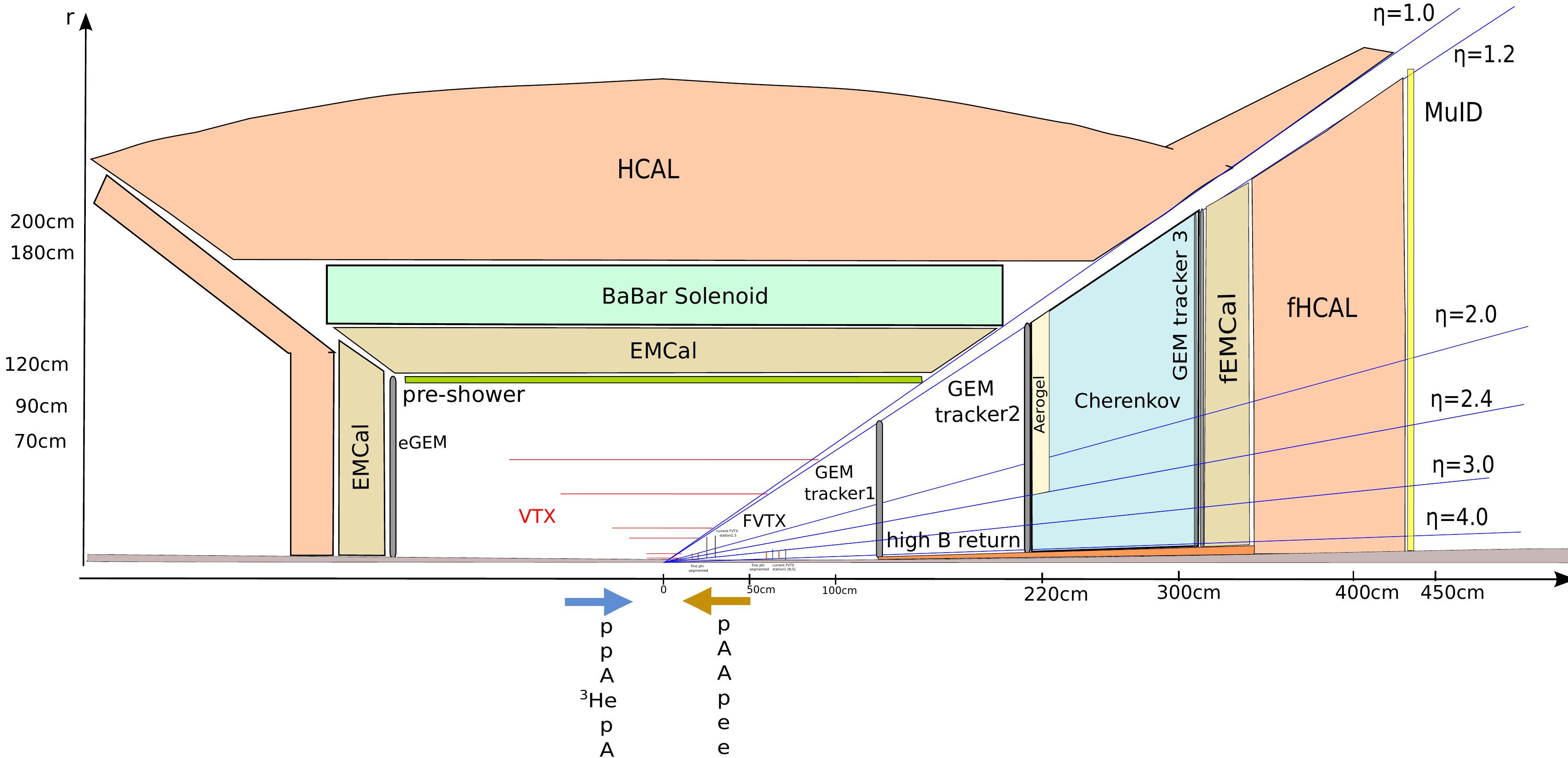}
\caption[]{sPHENIX concept with forward detector~\cite{sPHENIX2013}.
\label{fig:sPHENIX}}%\vspace*{-0.8pc}
\end{figure}

For PHENIX, the future plans are built around a new detector, sPHENIX, based on a superconducting thin-coil solenoid surrounded by EM and Hadron calorimeters in order to measure jets as well as to improve on other hard-scattering measurements by taking advantage of high-rate calorimeter triggers and the much larger solid angle. Personally, I am fond of such a detector because the first thin coil superconducting solenoid detector at a collider was used in the CCOR experiment at CERN in 1977 (Fig.~\ref{fig:solenoids}a) which made some of the major original hard-scattering measurements~\cite{RTbook,CMOR3jet}.  
This year, sPHENIX got a big boost by the acquisition of the BABAR solenoid magnet from SLAC (Fig.~\ref{fig:solenoids}b)  which became available when the B-factory in Italy was cancelled. The conceptual design of the new experiment is well underway~\cite{sPHENIX2013}, with mid-rapidity, forward and eRHIC capability (Fig.~\ref{fig:sPHENIX}). New collaborators are most welcome. 
      
\section*{References}
\bibliography{iopart-num-mjt}

\end{document}